\documentclass[8pt,aps,prb,nofootinbib,showpacs,twocolumn,latexsym]{revtex4}
\usepackage{amssymb}

\usepackage{latexsym}
\usepackage[dvips]{graphicx}
\usepackage{amsmath}
\usepackage[mathcal]{eucal}

\begin{document}

\title{Chiral spin resonance and spin-Hall conductivity in the presence of
the electron-electron interactions}
\author{ A.~Shekhter,$\;$ M.~Khodas and A.M.~Finkel'stein }

\begin{abstract}
We discuss the electron spin resonance in two-dimensional electron gas at zero external
magnetic field. This spin-resonance is due to the transitions between the
electron states, which are split by the spin-orbit (SO) interaction, and is
termed as the chiral spin resonance (CSR). It can be excited by the in-plane
component of the electric field of microwave radiation.
We show that there exists an inherent relationship between the spin-Hall
conductivity and the CSR in a system with the SO interaction. Since in the
presence of the SO interaction spin is not conserved, the electron-electron
interaction renormalizes the spin-Hall conductivity as well as the frequency
of the CSR. The effects of the electron interaction in systems with the SO
interaction are analyzed both phenomenologically and microscopically.
\end{abstract}

\affiliation{Department of Condensed Matter Physics, the Weizmann Institute of Science,
Rehovot, 76100, Israel}
\maketitle

%\date{\today }

\section{Introduction}
\label{sec:introduction}
In systems with spin-orbit (SO) interactions the spin of electrons can be
coupled to an electric field, making it possible to manipulate electron
spins without applying magnetic fields. This is the main reason why the
properties of the electron gas in the presence of the SO interaction are in
the focus of the research in spintronics.\cite{DattaDas90}

In semiconductors with a zinc-blende or a wurtzite lattice the SO
interaction originates from the bulk-inversion asymmetry (BIA) of the
crystal structure\cite{Dresselhaus55,RashbaSheka59}, whereas the
structure inversion asymmetry (SIA) typical for heterostructures is another
source of the SO interaction\cite{Vasko79,BychkovRashba84} in two-dimensional electron gas (2DEG).

In the presence of the SO interaction\ the spin degeneracy of the electron
spectrum is lifted. In this context, the possibility of the existence of the
spin-Hall current mediated by the SO interaction has been discussed
recently.\cite{Hirsh99,Murakami2003,Sinova2004,Hu2003,Schliemann2003,Rashba2003} It is now widely accepted\cite{Inoue2004,Mishchenko2004,Khaetskii2004,Raimondi2004,Rashba2004} that in the static limit the disorder
suppresses the spin-Hall conductivity in the bulk of a macroscopic system.\cite{Wunderlich04} Therefore, to investigate the bulk effects related to
the SO interaction it is worthwhile to turn to the high-frequency phenomena
when $\omega \tau \gg 1$.

In this paper we study the ac spin-Hall conductivity in a 2DEG with the
Bychkov-Rashba SO interaction.\cite{BychkovRashba84}  We demonstrate that similar to the Hall conductivity, which
in the absence of the SO interaction is inherently related to the
cyclotron resonance, the spin-Hall conductivity is related to a specific for
SO systems version of the electron spin resonance (ESR) which has been
termed by Rashba as a ''combined resonance''.\cite{Rashba60,McCombe69,RashbaSheka91} The combined resonance occurs as a result of the transitions
between the electron states, which are split by the combined action of the
SO interaction and the Zeeman interaction induced by a static magnetic
field. In a 2DEG with the Bychkov-Rashba SO interaction\cite{BychkovRashba84}
the spin-split eigenstates (in the limit of zero magnetic field) are
characterized by their chirality. We will be interested in the particular
limit of the combined resonance when a static magnetic field is absent and
the resonance is due to the transitions between electron states with
different chirality. To underline the nature of this resonance, we use the term
''chiral spin resonance'' (CSR), which 
emphasizes that the discussed electron-spin resonance occurs between the
chiral states that are spin split by the SO interaction rather than by the
external magnetic field.

In the presence of the SO interaction the dynamics of the total current and
the total spin is affected by the electron-electron (\emph{e-e})
interaction. Consequently, the frequency and the width of the CSR as well as
the spin-Hall conductivity acquire renormalization corrections. We start
this paper by applying the Kohn's theorem\cite{Kohn1961} procedure to
analyze the transverse transport coefficients in systems with the SO
interaction. In systems with \emph{no} SO interactions it is easy to show
that the absence of the \emph{e-e} renormalization of the Hall coefficient $%
R_{H}$ at $\omega _{c}\tau \gg 1$ is a direct consequence of the Kohn's
theorem. We observe, however, that the SO and the \emph{e-e} interactions
are not compatible in a sense that the equations of motion for the current
operators can be closed when only one of these interactions is present.
Still, this approach proves to be useful in finding a relation between the
spin-Hall conductivity and the dynamic spin-susceptibility that holds in the
clean limit ($\omega \tau \gg 1$) even in the presence of both the SO and
the \emph{e-e} interactions.\cite{Dimitrova2004}

A discussion of the Hall and spin-Hall conductivities following the lines of
the Kohn's theorem argumentation is given in Sec.~\ref{sec:eeinteraction}. A calculation of the spin-Hall conductivity in the absence of the \emph{e-e} interaction using the equation of motion for the current operators is
given in Appendix~\ref{sec:Kohntheorem}. (A reader not familiar with the
spin-Hall conductivity is recommended to look at the calculations in the
Appendix~\ref{sec:Kohntheorem} before proceeding further.) In Sec.~\ref{sec:flphenomenology} we consider the renormalization effects in the dynamic
spin-susceptibility induced by the \emph{e-e} interaction within the
framework of the phenomenological Fermi-liquid theory. We find the spectrum
of the spin excitations in the SO system and, in particular, determine the
frequency of the spin resonance. Simultaneously, we calculate the effects of
the \emph{e-e} renormalization on the spin-Hall conductivity. This is how
the relationship between the spin-Hall conductivity and the CSR\ can be
established. In Sec.~\ref{sec:flmicroscopic} an alternative microscopic
Fermi-liquid analysis of the dynamic spin susceptibility is presented for a
justification of the both approaches. In Sec.~\ref{sec:resonancebroadening} we find the disorder-induced width of the CSR
including its \emph{e-e} renormalizations. In addition, the Fermi-liquid
renormalizations of the D'yakonov-Perel rate of the spin relaxation \cite{DyakonovP71} are obtained. In the end of this Section we discuss the
electron-dipole mechanism \cite{RashbaSheka91} of the excitation of the CSR.

Finally, in the concluding section~\ref{sec:Summary} we discuss the
perspectives of the experimental observation of the CSR, i.e., the combined
resonance in the vanishing magnetic field. To observe the CSR, the spin
splitting induced by the SO interaction should be sufficiently isotropic.
For the purpose of definiteness, the calculation has been performed for the
case of Bychkov-Rashba SO interaction (SIA). However, the results of this
analysis are applicable in various other situations. In Appendix~\ref%
{sec:BIA} we discuss the forms of the SO interaction due to the
lack of the inversion symmetry of the host crystal (BIA)
\cite{Dresselhaus55} corresponding to quantum wells grown in
different crystallographic directions. We demonstrate that there
is a duality transformation relating the linear terms in the SO
interaction originating from the SIA and BIA mechanisms. Because of 
this duality all the conclusions about the spin-Hall conductivity
and the electron-spin resonance found for the Bychkov-Rashba SO
interaction hold equally well for the linear terms originating
from the Dresselhaus SO interaction in the cases of [001]- and
[111]-grown quantum wells.

\section{transverse conductivities in the presence of the electron-electron
interaction}
\label{sec:eeinteraction}
Let us start with the application of the Kohn's theorem procedure to the
Hall conductivity (and the Hall coefficient) in a system without a SO
interaction. In the presence of a magnetic field $B$ with the corresponding
vector potential ${\mathbf{A}}$ a many-electron system is described by the
Hamiltonian
\begin{equation}
H=\sum_{i}\frac{1}{2m}{\big[}\mathbf{p}_{i}{-\frac{e}{c}\mathbf{A}(}\mathbf{r%
}_{i}{)\big]}^{2}+\frac{1}{2}\sum_{i\neq j}V_{e-e}(\mathbf{r}_{i}-\mathbf{r}%
_{j})\;.  \label{eq:hinteract}
\end{equation}%
In this paper $m$ denotes the band-structure mass of an electron in a
heterostructure in contrast to $m_{e}$ denoting the vacuum mass of the
electron. The current operator in the presence of the vector potential ${%
\mathbf{A}}$ is
\begin{equation}
\mathbf{J}=\sum_{i}\frac{1}{m}\big[\mathbf{p}_{i}-\frac{e}{c}\mathbf{A}({\mathbf{r}}_{i})\big]\,.  \label{eq:currenttotal}
\end{equation}%
It is convenient to introduce ''the angular-momentum components'' of the
current operator $\mathbf{J}$
\begin{equation}
J^{\pm }=J^{x}\pm iJ^{y}\;  \label{eq:currenttrans}
\end{equation}
with the commutator
\begin{equation}
\lbrack J^{+},J^{-}]=-2\frac{\omega _{c}}{m}\hat{N},
\label{eq:currentcommute}
\end{equation}
where $\omega _{c}=|e|B/mc$ is the frequency of the cyclotron resonance and $\hat{N}$ is an operator of the total number of particles in the system. The
Kohn's theorem states that the \emph{e-e} interaction does not change the
frequency of the cyclotron resonance.\cite{Kohn1961} The essence of the
theorem is the observation that the electron interaction does not affect the
equation of motion for the total current operator

\begin{equation}
-i\frac{\partial }{\partial t}J^{\pm }(t)=\pm \omega _{c}J^{\pm }(t)\;.
\label{eq:eqmotionJ}
\end{equation}%
The unique property of the operators $J^{\pm }$ is that they connect the
pairs of states $l,m$ with the energy difference $E_{l}-E_{m}=\pm \,\omega
_{c}\,$only. The closed equation (\ref{eq:eqmotionJ}) yields for the time
dependence of the total current operators
\begin{equation}
J^{\pm }(t)=e^{\pm i\omega _{c}t}J^{\pm }\;.  \label{eq:timemotionJ}
\end{equation}%
With the Kohn's result for the time dependence of the current operators we
are fully equipped for the calculation of the conductivity tensor. According
to the Kubo formula, the conductivity tensor in the presence of the \emph{%
e-e } interaction is given by
\begin{align}
\sigma _{+-}& =\frac{e^{2}}{\omega }\int_{0}^{\infty }dte^{i\omega t}%
\big\langle\big[J^{+}(t),\;J^{-}(0)\big]\big\rangle=  \notag \\
& =-2i\frac{ne^{2}}{m\omega }\frac{\omega _{c}}{\omega +\omega _{c}},
\end{align}%
where $\langle \cdots \rangle $ means quantum-mechanical as well as thermal
average and $n$ is the density of the electron gas. Finally, having in mind
that $\sigma _{-+}(\omega )=-\sigma _{+-}(-\omega ),$ $\sigma _{++}=\sigma
_{--}=0,$ and $\sigma _{xy}=-\sigma _{yx}=(1/4i)\,\big[\sigma _{-+}-\sigma
_{+-}\big],$ one gets for the transverse components of the conductivity
tensor the following result:
\begin{equation}
\sigma _{xy}=-\frac{ne^{2}}{m}\frac{\omega _{c}}{\omega _{c}^{2}-\omega ^{2}}%
\;.  \label{eq:ordHall}
\end{equation}%
Remarkably, the factor $n$ preserves here its physical meaning of the
density of the electron gas and does not acquire any renormalization
correction in the presence of the electron interaction because of the universal
form of the commutator (\ref{eq:currentcommute}). Together with the absence
of the renormalization corrections to the cyclotron frequency this leads to
an important consequence for the Hall coefficient $R_{H}=\rho _{xy}/B$.
Inverting the conductivity tensor one obtains in the dc limit $\omega
\rightarrow 0$,
\begin{equation}
\rho _{xy}=\frac{m}{ne^{2}}\omega _{c}\,;\;\;R_{H}=-1/nec\;.
\label{eq:rhallcoef}
\end{equation}%
Thus, the absence of the renormalization corrections to $R_{H}$ in the clean
limit, $\omega _{c}\tau \gg 1,$ is a direct consequence of the Kohn's
theorem. For the limit of a weak magnetic field, $\omega _{c}\tau \ll 1,$
the proof of the absence of the renormalization corrections to the Hall
coefficient of an interacting electron gas requires a considerable effort.%
\cite{KFHall2003}

Let us check the possibility to extend the Kohn's theorem to a 2DEG with the
Bychkov-Rashba SO interaction \cite{BychkovRashba84} originating from the
structure-inversion asymmetry of the heterojunction
\begin{equation}
H^{SO}=\sum \alpha \lbrack \boldsymbol{p}_{i}\times \boldsymbol{\ell}]\cdot %
\boldsymbol{\sigma}_{i}\;,  \label{eq:BRashbaSO}
\end{equation}%
where the unit vector $\boldsymbol{\ell}$ is perpendicular to the plane of
the 2DEG. In the presence of the SO interaction the current operator $%
\mathbf{J}$ contains a spin-dependent term
\begin{equation}
\mathbf{J}=\sum \left( \frac{\boldsymbol{p}_{i}}{m}+\alpha \lbrack %
\boldsymbol{\ell}\times \boldsymbol{\sigma}_{i}]\,\right) \equiv \mathbf{P}%
/m+2\alpha \lbrack \boldsymbol{\ell}\times \mathbf{S}]\,.  \label{eq:currentSO}
\end{equation}%
Here $\mathbf{P}$ and $\mathbf{S}$ are the operators of the total momentum
and spin, respectively. Since the \emph{e-e} interaction commutes with the
current operator the interaction drops out from the equation of motion for $%
\mathbf{J,}$ as it takes place in the Kohn's theorem. Still, the current
operator has a complicated dynamics due to the SO interaction. For example,
for the component $J^{y}$ one gets
\begin{equation}
idJ^{y}/dt=-2i\alpha ^{2}\sum p_{i}^{x}\sigma _{i}^{z}=-4im\alpha ^{2}%
\mathfrak{J}_{z}^{x}\,,  \label{eq:motionJ}
\end{equation}%
where $\mathfrak{J}_{z}^{x}$ is the $x$ component of $z$-spin current
operator
\begin{equation}
\mathfrak{J}_{z}^{x}=\frac{1}{2}\sum \frac{p_{i}^{x}}{m}\sigma _{i}^{z}\;.
\label{eq:spincurrenttotal}
\end{equation}%
An attempt to get a closed system of equations by supplementing Eq.~(\ref%
{eq:motionJ}) with the equation of motion for $\mathfrak{J}_{z}^{x}$ fails.
It happens in the following way: in the equation of motion for the total
current $J^{\alpha }$ the contributions from the \emph{e-e }interaction term
$V_{e-e}$ cancel pairwise: $\partial V_{e-e}(r_{i}-r_{j})/\partial
r_{i}+\partial V_{e-e}(r_{i}-r_{j})/\partial r_{j}=0.$ On the contrary, in
the equation of motion for the spin current $\mathfrak{J}_{z}^{x}$ each of
the derivatives is multiplied by a spin operator of different particles and, as a result, the \emph{e-e} interaction does not drop out: $\sigma
_{i}^{z}\partial V_{e-e}(r_{i}-r_{j})/\partial r_{i}+\sigma _{j}^{z}\partial
V_{e-e}(r_{i}-r_{j})/\partial r_{j}=(\sigma _{i}^{z}-\sigma
_{j}^{z})\partial V_{e-e}(r_{i}-r_{j})/\partial x_{i}\neq 0$.

The very fact that $V_{e-e}$ does not drop out from the equations of motion
indicates that in the presence of the SO interaction the dynamics of the
electron gas is affected by the \emph{e-e} interaction. In spite of this
complication, the Kohn's theorem approach is useful for proving the relation
between the spin-Hall conductivity and the dynamic spin susceptibility that
remains intact even in the presence of the \emph{e-e} interaction (see also
Ref.~\onlinecite{Dimitrova2004}). The spin-Hall conductivity $\varsigma _{xy}^{z}$
describes the response of the spin-$z$-component current in the $x$ direction $\mathfrak{J}_{z}^{x}$ to the electric field applied in the $y$ direction. It is determined by the Kubo formula as follows:
\begin{equation}
\varsigma _{xy}^{z}=\frac{e}{\omega }\int_{0}^{\infty }dt\,e^{i\omega
t}\langle \left[ \mathfrak{J}_{z}^{x},J^{y}(-t)\right] \rangle \,.
\label{eq:spinHalltotal}
\end{equation}%
To explore its relation with the spin susceptibility we eliminate$\mathfrak{%
\ J}_{z}^{x}$ in favor of $J^{y}$ with the use of Eq.~(\ref{eq:motionJ}).
Performing\ the time integration by parts one obtains
\begin{equation}
\varsigma _{xy}^{z}=\frac{ie}{4m\alpha ^{2}}\int_{0}^{\infty }dt\,e^{i\omega
t}\langle \left[ J^{y}(t),J^{y}(0)\right] \rangle \,.  \label{eq:JJcor}
\end{equation}%
In a translation-invariant system the total momentum $\mathbf{P}(t)$ is a
conserved quantity, the commutator $[\mathbf{P}(t),\mathbf{S}(0\mathbf{)}]=[%
\mathbf{P}(0),\mathbf{S}(0)]~=~0,$ and, therefore, the momentum operator $%
P^{y} $ drops out from Eq.~(\ref{eq:JJcor}). Finally, one gets
\begin{equation}
\varsigma _{xy}^{z}(\omega )=\frac{ie}{m}\int_{0}^{\infty }dt\,e^{i\omega
t}\langle \lbrack S^{x}(t),S^{x}(0)\rangle \,.  \label{eq:SScor}
\end{equation}%
Thus, there is a direct connection between $\varsigma _{xy}^{z}$ and the
dynamic (retarded) spin susceptibility

\begin{eqnarray}
\varsigma _{xy}^{z} &=&\frac{e}{m}\chi _{xx}(q=0,\omega )\;,
\label{eq:cond-suscept} \\
\chi _{xx}(q=0,\omega )&=&\frac{i}{4}\int_{0}^{\infty }dt\,e^{i\omega
t}\langle \sum \left[ \sigma _{i}^{x}(t),\sigma _{i}^{x}(0)\right] \rangle
\,.  \notag
\end{eqnarray}%
In the presence of the SO interaction $\chi _{xx}$ has a behavior that
differs radically from that when $\alpha =0$. In the absence of the SO
interaction the total spin is conserved. Therefore, $\chi _{xx}(q=0,\omega
)_{\alpha =0}=0$ and, consequently, the spin-Hall conductivity vanishes at $%
\alpha =0$, whereas at any finite $\alpha $ one gets $\chi _{xx}(q=0,\omega
)\neq 0$ and, consequently, $\varsigma _{xy}^{z}\neq 0$.

It is worth noting that Eq.~(\ref{eq:cond-suscept}) is valid only in the
absence of disorder. The correlation function $\chi _{xx,}$ by itself, is
insensitive to disorder as long as the elastic scattering rate is less than the spin splitting energy. However, the
relation between $\varsigma _{xy}^{z}$ and $\chi _{xx}$ is very subtle\cite{Inoue2004,Mishchenko2004,Khaetskii2004,Raimondi2004,Rashba2004}, because in the presence of disorder the momentum is not
conserved, and there is a competition between the spin and momentum
contributions to the current vertex (inter- and intrabranch contributions in
terminology of Ref.~\onlinecite{Rashba2004}). The involvement of the momentum part
of the current operator makes Eq.~(\ref{eq:cond-suscept}) unapplicable for
analyzing the static limit of the spin-Hall conductivity in the presence of
disorder or an external magnetic field. Still, Eq.~(\ref{eq:cond-suscept})
is valid when $\omega \gg \eta _{2}$ [see Eq.~(\ref{eq:reswidth}) and the discussion in  the end of 
Sec.~\ref{sec:resonancebroadening}] and will be used for the analysis of the CSR.

\section{Fermi-liquid analysis of spin correlation function in the presence of
SO interaction. The spin resonance}
\label{sec:flphenomenology}
It has been demonstrated above that in the presence of the SO interaction,
the dynamics of the total current is affected by the \emph{e-e} interaction.
As a consequence of this fact, the spin-Hall conductivity acquires
corrections, which we analyze now with the use of the methods of the
phenomenological Fermi-liquid theory. Since the calculation of the spin-Hall
conductivity reduces in the clean limit to determining the dynamic
spin-susceptibility, one can follow the derivation of the spin-waves
spectrum in the Fermi liquid in an external magnetic field (see Chap. 1,
\S 5 in Ref.~\onlinecite{Pitaevskii}. There is an important difference, however,
between the spin splitting induced by the external magnetic field and the SO
interaction. As a result of the SO interaction the spin of an electron feels
an ''individual'' magnetic field, which is directed perpendicular to the
momentum of the electron. For this reason, to analyze the spin dynamics in
the presence of the SO interaction it is convenient to introduce the chiral
basis with the rotated Pauli matrices $\mathbf{\tau }_{{\mathbf{p}}}^{\nu }=(%
\mathbf{a}_{\mathbf{p}}^{\nu }\cdot \boldsymbol{\sigma}),$ where $\mathbf{a}%
_{\mathbf{p}}^{\nu }=\{\mathbf{a}^{1},\mathbf{a}^{2},\mathbf{a}^{3}\}=\{-%
\boldsymbol{\ell},\hat{\mathbf{p}},\hat{\mathbf{p}}\times \boldsymbol{\ell}%
\} $ and $\hat{\mathbf{p}}$ stands for a unit vector in the direction of
momentum ${\mathbf{p}}$. [Here we consider the Bychkov-Rashba SO
interaction. Similar analysis can be done for the case of SO interaction
induced by BIA\cite{Dresselhaus55} (see Appendix~\ref{sec:BIA} for details).]

Since $\mathbf{a}_{\mathbf{p}}^{\nu }$ form an orthonormal basis, $\tau$ matrices have the same commutation relations as the Pauli matrices. In the
chiral basis, the free single-particle Hamiltonian acquires the diagonal form
\begin{equation}
H^{SO}=\frac{\mathbf{p}^{2}}{2m}+\alpha |\mathbf{p}|\tau _{\mathbf{p}}^{3}
\label{eq:tausplit}
\end{equation}%
with the energy spectrum split into two chiral branches
\begin{equation}
\epsilon _{p}^{\pm }=p^{2}/2m\pm \alpha p\,.  \label{eq:chirlbr}
\end{equation}

In the presence of the \emph{e-e} interaction the spin splitting induced by
the SO interaction is renormalized. It can be determined by a
self-consistent equation
\begin{equation}
\delta \hat{\epsilon}_{\mathbf{p}}^{SO}=\alpha \big[\mathbf{p}\times %
\boldsymbol{\ell}\big] \cdot \boldsymbol{\sigma}+\mathrm{Tr^{\prime }}\int d\Omega
^{\prime }\hat{f}_{\mathbf{pp^{\prime }}}\frac{\partial n}{\partial \epsilon
}\delta \hat{\epsilon}_{\mathbf{p}^{\prime }}^{SO}.  \label{eq:energy}
\end{equation}%
where $\hat{f}_{\mathbf{pp^{\prime }}}$ is the function introduced by Landau
to describe the effects of electron interaction in the Fermi liquid, and $\mathrm{Tr^{\prime}}$ denotes the trace with respect to the spin indices.
In Eq.~(\ref{eq:energy}) $\delta \hat{n}_{\mathbf{p}}^{SO}=\partial
n/\partial \epsilon \delta \hat{\epsilon}_{\mathbf{p}}^{SO}$ is the response
of the distribution function of the quasiparticles to the SO-interaction
term, while the integral term describes the modification of the
quasiparticle energy spectrum as a result of the change of the
quasiparticle distribution. Note that Eq.~(\ref{eq:energy}) is a matrix
equation in spin space, and we use for the function $\hat{f}_{\mathbf{%
pp^{\prime }}}$ the standard notation $\nu (\epsilon _{F})\hat{f}_{\mathbf{%
pp^{\prime }}}=F(\theta _{\mathbf{p}\mathbf{p}^{\prime }})+G(\theta _{%
\mathbf{p}\mathbf{p}^{\prime }})\overrightarrow{\boldsymbol{\sigma}}\cdot
\overrightarrow{\boldsymbol{\sigma}}^{\prime }$, where $\theta _{\mathbf{p}%
\mathbf{p}^{\prime }}$ is an angle between $\mathbf{p}$ and $\mathbf{%
p^{\prime }}$, and $\nu (\epsilon _{F})=m^{\ast }/\pi $ is the renormalized
density of states for both spin components in a 2DEG. (Naturally, only the
spin-dependent part of the Landau's function is important for the
phenomena related to the SO interaction.) To solve Eq.~(\ref{eq:energy}) one
should expand $G(\theta )$ in a series of $2D$-harmonics, $G(\theta
)=\sum_{m}G^{m}e^{im\theta },$ and exploit the following property of the
Pauli matrices: $\overrightarrow{\boldsymbol{\sigma}}\cdot \mathrm{Tr\,}(%
\overrightarrow{\boldsymbol{\sigma}}^{\prime }\,\tau _{\mathbf{p}}^{\nu
})=2\tau _{\mathbf{p}}^{\nu }$. As a result, the renormalized spin splitting
of the electron energy spectrum, $\delta \hat{\epsilon}_{\mathbf{p}%
}^{SO}\equiv \alpha ^{\ast }p_{F}\tau _{\mathbf{p}}^{3}$ , is determined by
the renormalized SO parameter $\alpha ^{\ast }=\alpha /(1+G^{1})$; see also
Ref.~\onlinecite{Chen1999}.

To find the dynamic spin susceptibility $\chi (q=0,\omega )$ we calculate a
response linear in the time-dependent \emph{in-plane} magnetic field $B_{x}e^{i\omega t}$. Consider the equation of motion of the density matrix $\delta \hat{n}$ in the Landau's Fermi liquid in the presence of the SO
interaction and the perturbation term $\delta \hat{\epsilon}^{B}$, which is
introduced by the magnetic field,
\begin{equation}
\delta \hat{\epsilon}^{B}=-g\mu _{B}(\sigma ^{x}/2)B_{x}e^{i\omega
t}=-\sigma ^{x}\mathfrak{F}e^{i\omega t}\,,  \label{eq:force}
\end{equation}%
where $\mu _{B}=e\hbar /2m_{e}c$ and the Lande-factor $g$ depends on the
semiconductor. [In GaAs $g=$ $-0.44$, whereas in $In_xGa_{1-x}As$
heterostructures the absolute value $\left| g\right| $ can be an order of
magnitude larger]. Since spin variables are involved, $\delta \hat{n}$ is a
matrix in spin space and its time evolution is given by the commutator
\begin{equation}
i\frac{\partial }{\partial t}\delta \hat{n}_{\mathbf{p}}=[\delta \hat{n}_{\mathbf{p}},\delta \hat{\epsilon}_{\mathbf{p}}].  \label{eq:kinetic}
\end{equation}
In our case $\delta \hat{\epsilon}_{\mathbf{p}}=\alpha p_{F}\tau _{\mathbf{p}
}^{3}+\delta \hat{\epsilon}^{B}+\mathrm{Tr^{\prime }}\int d\Omega ^{\prime}
\hat{f}_{\mathbf{pp^{\prime }}}\delta \hat{n}_{\mathbf{p}^{\prime }}$, where
the last term accounts for the effects of the Fermi liquid.
 
To find the response linear in the magnetic field, one has to consider the
case when the magnetic term is much smaller than the spin-orbit one, $\delta
\hat{\epsilon}^{B}\ll 2\alpha p_{F}$. In equation~(\ref{eq:kinetic}) the
static part of $\delta \hat{n}_{\mathbf{p}}$ induced by the SO interaction, $\delta \hat{n}_{\mathbf{p}}^{SO},$ should be separated from a time-dependent
part $\delta \hat{n}_{\mathbf{p}}^{B}\,$
\begin{equation}
\delta \hat{n}_{\mathbf{p}}=\delta \hat{n}_{\mathbf{p}}^{SO}+\delta \hat{n}_{\mathbf{p}}^{B}.  \label{eq:separate}
\end{equation}%
After this separation Eq.~(\ref{eq:kinetic}) can be linearized with respect
to $\delta \hat{\epsilon}^{B}$ and $\delta \hat{n}_{\mathbf{p}}^{B}$
\begin{align}
i\frac{\partial }{\partial t}\hat{u}_{\mathbf{p}}=& -[\delta \hat{\epsilon}_{\mathbf{p}}^{SO}\;,\;\hat{u}_{\mathbf{p}}+\nu (\epsilon _{F})\frac{1}{2}
\mathrm{Tr^{\prime }}\int d\Omega ^{\prime }\hat{f}_{\mathbf{pp^{\prime }}}
\hat{u}_{\mathbf{p}^{\prime }}]  \notag \\
+& [\delta \hat{\epsilon}_{\mathbf{p}}^{SO},\,\sigma _{x}]\;\mathfrak{F}
e^{i\omega t}\;.  \label{eq:linearized}
\end{align}%
Here we rewrite $\delta \hat{n}_{\mathbf{p}}^{B}$ in terms of the
displacement function $\hat{u}_{\mathbf{p}}$, describing the deformation of
the Fermi surface, $\delta \hat{n}_{\mathbf{p}}^{B}=(\partial n/\partial
\epsilon )\,\,\hat{u}_{\mathbf{p}}$ (note that $\hat{u}_{\mathbf{p}}$
depends on the direction of the vector $\mathbf{p}$ and is a matrix in spin
space). In Eq.~(\ref{eq:linearized}) the static part $\delta \hat{n}_{\mathbf{p}}^{SO}$ has been absorbed by $\delta \hat{\epsilon}_{\mathbf{p}}^{SO}$ giving the renormalized spin-splitting energy $\Delta $
\begin{equation}
\delta \hat{\epsilon}_{\mathbf{p}}^{SO}=\alpha ^{\ast }p_{F}\tau _{\mathbf{p}}^{3}=\frac{1}{2}\Delta \tau _{\mathbf{p}}^{3}\,.  \label{eq:SOdelta}
\end{equation}%
With the use of $\sigma _{x}=(p_{x}\tau _{\mathbf{p}}^{2}+p_{y}\tau _{\mathbf{p}}^{3})/p$ the ''driving-force'' term in the above equation can be
rewritten as
\begin{equation}
\lbrack \delta \hat{\epsilon}_{\mathbf{p}}^{SO},\,\sigma _{x}]\,\mathfrak{F}\,e^{i\omega t}=-i\Delta \tau ^{1}\,\,\frac{p_{x}}{p}\,\mathfrak{F}\,e^{i\omega t}.  \label{eq:drivingforce}
\end{equation}

To solve Eq.~(\ref{eq:linearized}), one have to represent the matrix $\hat{u}_{\mathbf{p}}$ in terms of $\tau$ matrices: $\hat{u}_{\mathbf{p}}=u_{1}(\theta _{\mathbf{p}})\tau ^{1}+u_{2}(\theta _{\mathbf{p}})\tau _{\mathbf{p}}^{2}+u_{3}(\theta _{\mathbf{p}})\tau _{\mathbf{p}}^{3}$, where $\theta _{\mathbf{p}}$ denotes the direction of the vector$~\mathbf{p}$. The
coefficients $u_{i}(\theta _{\mathbf{p}})$ are determined by a system of
equations ($\mathbf{\theta _{\mathbf{p}\mathbf{p}^{\prime }}\equiv \theta _{\mathbf{p}}-\theta _{\mathbf{p}^{\prime }}}$)
\begin{subequations}
\begin{align}
\Delta ^{-1}\frac{du_{1}(\theta _{\mathbf{p}})}{dt}=& u_{2}(\theta _{\mathbf{p}})+\int d\theta _{\mathbf{p}^{\prime }}G(\theta _{\mathbf{p}\mathbf{p}^{\prime }})\cos \theta _{\mathbf{p}\mathbf{p}^{\prime }}u_{2}(\theta _{\mathbf{p}^{\prime }})  \notag \\
&\quad\quad\quad-\mathfrak{F}e^{i\omega t}\cos \theta _{\mathbf{p}}\;,
\label{eq:u-eqa} \\
\Delta ^{-1}\frac{du_{2}(\theta _{\mathbf{p}})}{dt}=& -u_{1}(\theta _{\mathbf{p}})-\int d\theta _{\mathbf{p}^{\prime }}G(\theta _{\mathbf{p}\mathbf{p}^{\prime }})u_{1}(\theta _{\mathbf{p}^{\prime }})\;,
\label{eq:u-eqb} \\
\frac{du_{3}(\theta _{\mathbf{p}})}{dt} =&0\;.  \label{eq:u-eqc}
\end{align}
In the transition from Eq.~(\ref{eq:linearized}) to Eqs.~(\ref{eq:u-eqa})
and (\ref{eq:u-eqb}) it has been used that $\overrightarrow{%
\boldsymbol{\sigma}}\cdot \mathrm{Tr}\left( \overrightarrow{%
\boldsymbol{\sigma}}\tau _{\mathbf{p}^{\prime }}^{\nu }\right) =2\tau _{%
\mathbf{p}^{\prime }}^{\nu }$ and that the commutator $[\tau _{\mathbf{p}%
}^{3},\tau _{\mathbf{p}^{\prime }}^{2}]=-2i\tau ^{1}\cos \theta _{\mathbf{p}%
\mathbf{p}^{\prime }}.$

Since the function $\hat{f}_{\mathbf{pp^{\prime }}}$ depends on the
directions of vectors $\mathbf{p}$ and $\mathbf{p^{\prime }}$ through $\cos
\theta _{\mathbf{p}\mathbf{p}^{\prime }}$ only, Eqs.~(\ref{eq:u-eqa})-(\ref{eq:u-eqc}) can be solved by expanding $u_{i}(\theta )$ in $2D$ harmonics, $u_{i}(\theta )=\sum_{m}u_{i}^{m}e^{im\theta }$
\end{subequations}
\begin{subequations}
\begin{align}
\Delta ^{-1}\;\frac{du_{1}^{m}}{dt}=& u_{2}^{m}\;\left[ 1+\frac{1}{2}%
(G^{m+1}+G^{m-1})\right]   \label{eq:harma} \\
& \quad -\frac{1}{2}(\delta _{m,1}+\delta _{m,-1})\,\mathfrak{F}e^{i\omega
t}\;\;\;,  \notag \\
\Delta ^{-1}\;\frac{du_{2}^{m}}{dt}=& -u_{1}^{m}\;(1+G^{m})\;.
\label{eq:harmb}
\end{align}%
It has been used here that $G(\theta )\cos \theta \rightarrow
(1/2)(G^{m+1}+G^{m-1})$ and that the harmonics coefficients are even in $m$,
$G^{m}=G^{-m}$ (this is because the function $\hat{f}_{\mathbf{pp^{\prime }}}
$ is even in $\theta _{\mathbf{p}\mathbf{p}^{\prime }}$). After the
time-Fourier transform one gets the frequencies of the spin waves
\end{subequations}
\begin{equation}
\omega _{m}^{2}(q=0)=\Delta ^{2}(1+G^{m})\Big[1+\frac{1}{2}(G^{m+1}+G^{m-1})\Big].  \label{eq:resonances}
\end{equation}%
In the absence of the \emph{e-e} interaction $\omega _{m}$ does not depend
on $m$ as each spin precess independently in the individual field
induced by the SO interaction with the same frequency. The \emph{e-e}
interaction couples the precession motion of different spins, thereby
lifting the degeneracy of the precession by renormalizing the frequency. As
a result one gets a set of the spin-wave excitations corresponding to
different $2D$-harmonics. Unlike the noninteracting case where the spin
precession is circular, in the presence of the \emph{e-e} interaction, the
precession is elliptical.

When an electromagnetic field is applied to the electron gas the CSR can be
excited. The only harmonics activated by the in-plane field are those with $%
m=\pm 1$
\begin{subequations}
\begin{align}
{u}_{1}^{\pm 1}=& -i\frac{\omega \Delta }{2(\omega _{1}^{2}-\omega ^{2})}%
\mathfrak{F}e^{i\omega t}\,,  \label{eq:firstharma} \\
{u}_{2}^{\pm 1}=& (1+G^{1})\frac{\Delta ^{2}\ }{2(\omega _{1}^{2}-\omega
^{2})}\mathfrak{F}e^{i\omega t}\,,  \label{eq:firstharmb}
\end{align}%
and, therefore, the CSR frequency $\omega ^{CSR}$ is determined by $\omega_{1}$
\end{subequations}
\begin{equation}
\omega ^{CSR}=\Delta \Big\{(1+G^{1})\big[1+\frac{1}{2}(G^{0}+G^{2})\big]\Big\}^{1/2}\,.
\label{eq:resfreq}
\end{equation}%
Unlike the ESR in the absence of the SO where the resonance frequency is
not renormalized, $\omega ^{CSR}$ is renormalized by the \emph{e-e}
interaction. This is quite natural as the ESR analog of the Kohn's theorem%
\cite{Kohn1961} does not hold in the presence of the SO interaction.

To find the spin-spin correlation function $\chi _{xx}$ we calculate $S_{x}=\frac{1}{2}\nu(\epsilon_{F})\mathrm{Tr}\int\hat{u}_{\mathbf{p}}(\sigma^{x}/2)(d\theta/2\pi)$ as a response of the electron gas to the
magnetic field $B_{x}$. Only $\tau _{\mathbf{p}}^{2}$-component of $\hat{u}_{%
\mathbf{p}}$ contributes and, therefore,
\begin{align}
S_{x}=& \frac{1}{4}\nu (\epsilon _{F})\sum_{m}u_{2}^{m}[\delta _{m,1}+\delta
_{m,-1}]=  \notag \\
=& \nu (\epsilon _{F})(1+G^{1})\frac{\Delta ^{2}}{8[(\omega
^{CSR})^{2}-\omega ^{2}]}(g\mu _{B})B_{x}\;.
\end{align}%
Noting that $\chi_{xx}$ is equal to $S_{x}/(g\mu _{B}B_{x})$, one obtains
in the limit of small $\omega \ll \omega ^{CSR}\sim \Delta $ that
\begin{equation}
\chi _{xx}(q=0,\omega \rightarrow 0)_{\alpha \neq 0}=\frac{1}{8}\frac{\nu
(\epsilon _{F})}{1+\frac{1}{2}(G^{0}+G^{2})}\;,  \label{eq:spinsusc}
\end{equation}%
and correspondingly in the absence of disorder the renormalized value of the
spin-Hall conductivity in the limit of small frequency (see the discussion
in the end of Section~\ref{sec:eeinteraction}) is equal to
\begin{equation}
\varsigma _{xy}^{z}=\frac{e}{8\pi }\frac{1}{1+\frac{1}{2}(G^{0}+G^{2})}\;%
\frac{m^{\ast }}{m}\;.  \label{eq:final}
\end{equation}%
The angular structure of the corrections to $\varsigma _{xy}^{z}$ calculated
in Ref.~\onlinecite{Dimitrova2004} to the lowest order in the \emph{e-e}
interaction are in agreement with this result.

In the above consideration we find the dynamic spin susceptibility by calculating
the response to a magnetic field $Be^{i\omega t}$ coupled to the magnetic
moments of electrons via the Zeeman interaction. Actually, as a mechanism of
the excitation of the CSR this type of coupling is very ineffective. In the
presence of the SO interaction the electromagnetic field $\mathbf{A}$ can
excite the spin-flip transitions  much more effectively by coupling through the electric-dipole
interaction $-(e/c)\mathbf{AJ}$. (For the
electric-dipole excited spin resonance see Ref.~\onlinecite{RashbaSheka91}). The
relative effectiveness of the two mechanisms is of the order of the ratio of
the Compton length to the electron wavelength: $(\lambdabar/\lambda
)^{2}\sim 10^{-9}-10^{-8}.$ We postpone the discussion of excitation of the
CSR as well as of the width of the resonance to Sec.~\ref{sec:resonancebroadening}.

Finally, a further comment is in order. The above calculation demonstrates
an inherent relationship between the spin-Hall conductivity and the CSR. The
same correlation function, $\chi _{xx}(q=0,\omega )_{\alpha \neq 0}$~,
describes the resonance and determines the value of $\varsigma _{xy}^{z}$,
including its static limit. Actually, the existence of a relationship
between a transverse conductivity and a corresponding resonance is generic.
In clean systems, in the absence of dissipation, the longitudinal
conductivity $\sigma _{xx}(\omega )$ vanishes when the frequency $\omega $
is in the range $1/\tau \ll \omega \ll \Delta E$ as at such frequencies the
dipole transitions with the energy $\Delta E$ cannot be excited. Unlike the
longitudinal, the Hall conductivity, as well as the spin-Hall one, are
related not to the real transitions but to the virtual. This leads to the
generic relationship between the transverse conductivities and the
corresponding resonance; see also Sec.~\ref{sec:eeinteraction},
where the connection
between the Hall conductivity and the cyclotron resonance has been
demonstrated.

\section{ Spin-susceptibility in the presence of the SO interaction: microscopic calculation}
\label{sec:flmicroscopic}
In this section we develop a microscopic derivation of the dynamic
spin susceptibility as an alternative to the phenomenological description
presented in Sec.~\ref{sec:flphenomenology}. As a whole, we follow the scheme
elaborated for the microscopic derivation of the dynamic spin susceptibility
by one of us in Ref.~\onlinecite{AF1984}.
\begin{figure}[h]
\centerline{
    \includegraphics[width=0.4\textwidth]{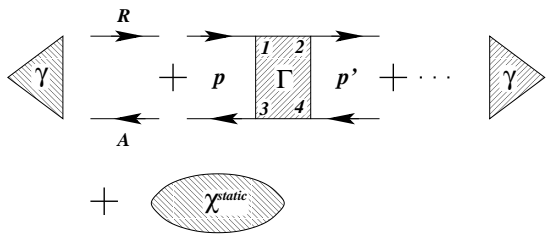}}
\caption{Spin-density correlation function}
\label{fig:flladder}
\end{figure}

Let us discuss the ladder diagrams for the spin-density correlation function
presented in Fig.~\ref{fig:flladder}. We choose to work with the
amplitudes known in the Fermi-liquid theory as $\Gamma ^{k}$. This approach
has the following reasoning. A two-particle vertex function $\Gamma (\omega ,%
\mathbf{k)}$ includes an irreducible part, the contributions from the
incoherent scattering, and, most importantly, the contributions from a
multiple rescattering of electron-hole quasiparticle pairs. Apart from
small corrections $\sim $ $\alpha /v_{F}\ll 1$, neither the irreducible
part of the vertex function nor the contributions from the incoherent
scattering are sensitive to a small modification of the electron spectrum
because they accumulate their values far from the Fermi surface. On the
contrary, the rescattering of a quasiparticle pair requires certain care.
The reason is that the contribution from a cross section with an electron-hole
pair as an intermediate state is equal to a singular combination $\mathbf{v}%
_{F}\mathbf{k/(}i\omega _{n}-\mathbf{v}_{F}\mathbf{k}),$ where $\mathbf{v}%
_{F}\mathbf{k}$ originates from $\delta \epsilon _{k}(\mathbf{p)}=\epsilon (%
\mathbf{p}+\mathbf{k})-\epsilon (\mathbf{p})$ (see Chap.~2, \S~17 in Ref.~\onlinecite{Pitaevskii}). Index $k$ in $\Gamma ^{k}$ means that in $\Gamma (\omega ,%
\mathbf{k)}$ the contributions of such cross sections are taken as follows:
one first takes $\omega =0$ and only afterward takes the limit $%
k\rightarrow 0$. In the presence of the SO interaction the energy difference
$\delta \epsilon _{k}(\mathbf{p)}$ acquires a gap when the two
quasiparticles have different chirality. The order of limits corresponding
to $\Gamma ^{k}$ makes this amplitude to be not sensitive to a gap in the
energy spectrum of the quasiparticles. Indeed, in the considered order of
limits the combination $\delta\epsilon_{\mathbf{k}}(\mathbf{p})/[i\omega_{n}-\delta\epsilon_{\mathbf{k}}(\mathbf{p})]$ is equal to~$-1$ for any
energy spectrum of electrons. Altogether this argumentation\cite{AF1983}
leads to the conclusion that the values of the static amplitudes $\Gamma^{k}$ are not modified by a SO interaction apart from small corrections $\sim \alpha /v_{F}\ll 1$. This feature of the amplitude $\Gamma^{k}$
makes it particularly convenient for the purposes of the microscopic
analysis. Diagrammatically the amplitude $\Gamma ^{k}$ can be defined as a
two-particle amplitude irreducible with respect to a $RA$ section (by the $RA$ section we understand a product of the two Green's functions when one of
them is retarded, while the other one is advanced). With the use of $\Gamma^{k}$, the ladder diagrams for the two-particle correlation functions are
rearranged in such a way that the blocks of the combination 
$i\omega _{n}/[i\omega_{n}-\delta\epsilon_{\mathbf{k}}(\mathbf{p})]$
rather than 
$\delta\epsilon_{\mathbf{k}}(\mathbf{p})/[i\omega_{n}-\delta\epsilon_{\mathbf{k}}(\mathbf{p})]$
stand separated by amplitudes $\Gamma ^{k}$.

Depending on the spin structure the two-particle amplitude can be split into
two parts
\begin{align}
\frac{\nu (\epsilon _{F})}{2}a^{2}\hat{{\Gamma }}{_{1}^{k}}\,_{\alpha
_{3}\,\alpha _{4}}^{\alpha _{1}\,\alpha _{2}}(\mathbf{p},\mathbf{p}^{\prime
})=& \Gamma _{1}(\mathbf{p},\mathbf{p}^{\prime })\delta _{\alpha _{1},\alpha
_{3}}\delta _{\alpha _{2},\alpha _{4}}  \notag \\
\frac{\nu (\epsilon _{F})}{2}a^{2}\hat{{\Gamma }}{_{2}^{k}}\,_{\alpha
_{3}\,\alpha _{4}}^{\alpha _{1}\,\alpha _{2}}(\mathbf{p},\mathbf{p}^{\prime
})=& -\Gamma _{2}(\mathbf{p},\mathbf{p}^{\prime })\delta _{\alpha
_{1},\alpha _{2}}\delta _{\alpha _{3},\alpha _{4}}\;.  \label{eq:ampl}
\end{align}%
Here matrices $\hat{{\Gamma }}{_{1,2}^{k}}$ denote the spin-dependent
amplitudes, while the dimensionless function $\Gamma _{1,2}$ determine the
parameters of the Fermi-liquid theory (in $\Gamma _{1,2}$ the index $k$ is
omitted). The minus sign in the amplitude $\Gamma _{2}$ is due to the
anticommutation of the fermionic operators. The factor $a$ appears in a
standard way because it describes the weight of the quasiparticle part in 
the Green's function.\cite{Pitaevskii} For electrons at the Fermi energy the
functions $\Gamma _{1,2}(\mathbf{p},\mathbf{p}^{\prime })= \Gamma
_{1,2}(\theta _{\mathbf{pp}^{\prime }})$ depend on the scattering angle $\theta _{\mathbf{p}\mathbf{p}^{\prime }}$ only. The coefficients of the
expansion of $\Gamma _{1,2}(\theta _{\mathbf{pp}^{\prime }})$ in angular
harmonics are used as the parameters of the Fermi-liquid theory. In $2D$
they are defined as follows:
\begin{equation}
\Gamma _{1,2}^{m}=\int \frac{d\theta }{2\pi }\Gamma _{1,2}(\theta )\exp
(-im\theta )\,.
\end{equation}

Now we turn to triangle vertices $\gamma .$ Like $\Gamma ^{k}$,
the vertex $ \gamma $ in Fig.~\ref{fig:flladder} is a dressed
vertex irreducible with respect to $RA$ sections (i.e., it extends
from an external vertex to the first $RA\;$section). The
arguments concerning insensitivity of the static limit of the
two-particle vertex functions to the SO interaction remains also
valid (with an accuracy $\sim $ $\alpha /v_{F}\ll 1$) for the
renormalized ''triangle'' vertex $\gamma $. Since we are
interested in the spin-density correlation function, we consider
the case when the external vertices contain a spin operator
$\sigma _{x}/2$ (such a vertex is denoted below as $\hat{{\gamma
}}^{\sigma _{x}}$). Due to the Fermi-liquid corrections the vertex
$\hat{{\gamma }}^{\sigma _{x}}$ acquires the renormalization
factor $(1+\Gamma _{2}^{0}),$ where $\Gamma _{2}^{0}$ is zero
harmonics of the interaction amplitude $\Gamma
_{2}(\mathbf{p},\mathbf{p} ^{\prime })$. The last contribution in
Fig.~\ref{fig:flladder} to be commented on is the static spin
susceptibility $\chi _{xx}(\omega =0)$ (this correlation function
does not contain any $RA\;$sections). According to the same
argumentation it is equal to $(1/4)\nu (\epsilon _{F})(1+\Gamma
_{2}^{0}).$

The singular part of the matrix Green's function in the presence of the
Bychkov-Rashba SO interaction is
\begin{equation}
\hat{\mathcal{G}}(i\epsilon ,\mathbf{p})=\sum_{\zeta =\pm 1}|\zeta \mathbf{p}
\rangle \frac{a}{i\epsilon -\epsilon _{p}^{\zeta }+\mu }\langle \zeta
\mathbf{p}|  \label{eq:Green}
\end{equation}
where the residue $a$ is a weight of the quasiparticle part in the Green's
function. [In what follows the singular parts of the Green's functions will
be used without the factor $a$. This is the reason for attaching $a^{2}$ to
the matrices $\hat{{\Gamma }}{_{1,2}^{k}}$ in the relations (\ref{eq:ampl}).
With the use of the effective mass $m^{\ast \text{ }}$ and a proper
redefinition of triangle vertices $\gamma $ the explicit dependence on the
residue $a$ drops out from the Fermi-liquid calculations.\cite{Pitaevskii}]
The direct product of spinors $|\zeta \mathbf{p}\rangle \langle \zeta
\mathbf{p}|$ in the Green's function $\hat{\mathcal{G}}(i\epsilon ,\mathbf{p}
)$ is the projector onto the chiral states with the eigenenergies $\epsilon
_{p}^{\zeta }=p^{2}/2m^{\ast }+\zeta \Delta /2$; here and in what follows the chiral state index $\zeta=\pm 1$.  The eigenspinors in Eq.~(\ref{eq:Green}) can be found from the eigenvalue problem for $\tau _{\mathbf{p}}^{3}$-matrix
\begin{align}
\tau _{\mathbf{p}}^{3}\,|\zeta \mathbf{p}\rangle =& \zeta \,|\zeta \mathbf{p}
\rangle ,  \notag \\
|\zeta \mathbf{p}\rangle =& \frac{1}{\sqrt{2}}\left[
\begin{array}{c}
i\zeta e^{-i\vartheta _{\mathbf{p}}} \\
1%
\end{array}
\right]  \label{eq:spinors}
\end{align}
where $e^{\pm i\vartheta _{\mathbf{p}}}=(p_{x}\pm ip_{y})/p$.

To conduct the calculation in the chiral basis, the spinors will be
transferred from the Green's functions to the interaction amplitudes and to
the vertices $\hat{{\gamma }}^{\sigma _{x}}$ (see Fig.~\ref{fig:flladder}). As a result, one gets for the matrices $\hat{{\Gamma }}{_{1,2}}$
\begin{align}
\frac{\nu (\epsilon _{F})}{2}a^{2}\hat{{\Gamma }}{_{1}^{k}\,}_{\zeta
_{3}\,\zeta _{4}}^{\zeta _{1}\,\zeta _{2}}(\mathbf{p},\mathbf{p}^{\prime
})=& \;\;\;\;\Gamma _{1}(\mathbf{p},\mathbf{p}^{\prime })\;\langle \zeta
_{3} \mathbf{p}|\zeta _{1}\mathbf{p}\rangle \;\langle \zeta _{2}\mathbf{p}
^{\prime }|\zeta _{4}\mathbf{p}^{\prime }\rangle \;, \notag \\
\frac{\nu (\epsilon _{F})}{2}a^{2}\hat{{\Gamma }}{_{2}^{k}\,}_{\zeta
_{3}\,\zeta _{4}}^{\zeta _{1}\,\zeta _{2}}(\mathbf{p},\mathbf{p}^{\prime
})=& -\Gamma _{2}(\mathbf{p},\mathbf{p}^{\prime })\;\langle \zeta _{2}
\mathbf{p}^{\prime }|\zeta _{1}\mathbf{p}\rangle \;\langle \zeta _{3}\mathbf{%
\ p}|\zeta _{4}\mathbf{p}^{\prime }\rangle \;,  \label{eq:gammachiral}
\end{align}
and, similarly, for the vertices $\hat{{\gamma }}^{\sigma _{x}}$
\begin{align}
a\,\hat{{\gamma }}^{\sigma _{x}}\,_{\zeta _{2}}^{\zeta _{1}}(\mathbf{p})=&
(1+\Gamma _{2}^{0})\,\langle \zeta _{1}\mathbf{p}|\frac{\sigma _{x}}{2}
|\zeta _{2}\mathbf{p}\rangle \;,  \notag \\
a\,\,_{\zeta _{2}}^{\zeta _{1}}\hat{{\gamma }}^{\sigma _{x}}(\mathbf{p})=&
(1+\Gamma _{2}^{0})\,\langle \zeta _{2}\mathbf{p}|\frac{\sigma _{x}}{2}
|\zeta _{1}\mathbf{p}\rangle \;,  \label{eq:vertexchiral}
\end{align}
where the first and second lines correspond to the left and right triangle
vertices in Fig.~\ref{fig:flladder}.

The matrix elements appearing in Eqs.~(\ref{eq:gammachiral}) and (\ref%
{eq:vertexchiral}) can be easily found:
\begin{align}
\langle \zeta \mathbf{p}|\zeta^{\prime} \mathbf{p}^{\prime }\rangle =& \frac{\zeta
\zeta ^{\prime }e^{i(\vartheta _{\mathbf{p}}-\vartheta _{\mathbf{p}^{\prime
}})}+1}{2}  \;,\notag \\
\langle \zeta \mathbf{p}|\sigma _{x}|\zeta^{\prime} \mathbf{p}^{\prime }\rangle =&
\frac{\zeta e^{i\vartheta _{\mathbf{p}}}-\zeta ^{\prime }e^{-i\vartheta _{%
\mathbf{p}^{\prime }}}}{2i}\,.  \label{eq:matrixelements}
\end{align}%
Then, for the matrix $\hat{\Gamma}_{1}$ one has
\begin{equation}
\frac{\nu (\epsilon _{F})}{2}a^{2}\hat{{\Gamma }}{_{1}^{k}\,}_{\zeta
_{3}\,\zeta _{4}}^{\zeta _{1}\,\zeta _{2}}(\mathbf{p},\mathbf{p}^{\prime })=%
\frac{1}{4}\Gamma _{1}(\vartheta _{\mathbf{\ pp^{\prime }}})(1+\zeta
_{1}\zeta _{3})(1+\zeta _{2}\zeta _{4})\,,  \label{eq:gammaone}
\end{equation}%
and for the matrix $\hat{\Gamma}_{2}$, which is of special importance since
it controls the dynamics of spins, one gets ($\vartheta_{\mathbf{pp^{\prime}}}=\vartheta_{\mathbf{p}}-\vartheta_{\mathbf{p}^{\prime}}$)
\begin{align}
\frac{\nu (\epsilon _{F})}{2}a^{2}\hat{{\Gamma }}{_{2}^{k}\,}_{\zeta
_{3}\,\zeta _{4}}^{\zeta _{1}\,\zeta _{2}}(\mathbf{p},\mathbf{p}^{\prime })=&-\frac{1}{4}\Gamma _{2}(\vartheta _{\mathbf{pp^{\prime }}})[1+\zeta
_{1}\zeta _{2}\zeta _{3}\zeta _{4}+\zeta _{1}\zeta _{2}e^{-i\vartheta _{%
\mathbf{pp^{\prime }}}}\notag\\
+&\zeta _{3}\zeta _{4}e^{i\vartheta _{\mathbf{pp^{\prime }}}}]\;.  \label{eq:gammatwo} 
\end{align}%
Note that there appears an additional angular dependence because of the
factors $e^{\pm i\vartheta _{\mathbf{pp^{\prime }}}}$and therefore in the
expansion in a series of $2D$-harmonics one should take into consideration
that $\Gamma (\vartheta )e^{\pm i\vartheta }\rightarrow \Gamma ^{m\mp 1}$ .

It will be convenient to represent the matrices $\hat{{\Gamma }}$ in the chiral basis by $4\times 4$ matrices $\hat{\Gamma}_{ij}$. For that we choose the following
convention. The first index $i$ represents the left pair of indices $\left(
\,_{\zeta _{3}}^{\zeta _{1}}\right) $ in the order $\left( \,_{+}^{+}\right)
,\left( \,_{-}^{-}\right) ,\left( \,_{-}^{+}\right) ,\left(
\,_{+}^{-}\right) $ , while the second index $j$ represents the right pair
of indices $\left( \,_{\zeta _{4}}^{\zeta _{2}}\right) $ in the same order.
Finally, after the expansion in $2D$-harmonics, the explicit block-form
expressions for the matrices $\hat{\Gamma}_{ij}$ look as follows ($\Sigma
_{x}=\left|
\begin{array}{cc}
0 & 1 \\
1 & 0%
\end{array}
\right| )$ :
\begin{align}
\frac{\nu (\epsilon _{F})}{2}a^{2}\hat{{\Gamma }}_{2}^{k}\,(m)_{ij}& =\quad
\quad \quad \quad \quad \quad \quad \quad  \label{eq:gammamatrix2} \\
-\frac{\Gamma _{2}^{m}}{2}\left|
\begin{array}{cc}
1+\Sigma _{x} & 0 \\
0 & 1+\Sigma _{x}%
\end{array}%
\right| -& \frac{\Gamma _{2}^{m+1}}{4}\left|
\begin{array}{cc}
1-\Sigma _{x} & 1-\Sigma _{x} \\
1-\Sigma _{x} & 1-\Sigma _{x}%
\end{array}%
\right|  \notag \\
-& \frac{\Gamma _{2}^{m-1}}{4}\left|
\begin{array}{rr}
1-\Sigma _{x} & -1+\Sigma _{x} \\
-1+\Sigma _{x} & 1-\Sigma _{x}%
\end{array}%
\right| \;,  \notag
\end{align}%
and
\begin{equation}
\frac{\nu (\epsilon _{F})}{2}a^{2}\hat{{\Gamma }}_{1}^{k}\,(m)_{ij}=\Gamma
_{1}^{m}\left|
\begin{array}{cc}
1+\Sigma _{x} & 0 \\
0 & 0%
\end{array}%
\right| \,.  \label{eq:gammamatrix1}
\end{equation}% 
Note, that one can combine $\hat{{\Gamma }}{_{1}}$ with the top-left block
in the first term of $\hat{{\Gamma }}{_{2}}$ to create the amplitude $%
(\Gamma _{1}-\Gamma _{2}/2)(1+\Sigma _{x})$, which controls the singlet
channel of the electron-hole excitations (charge-density excitations). Since
the spin dynamics is controlled by $(1-\Sigma _{x})$ and $(1+\Sigma_{x})(1-\Sigma _{x})=0$, the singlet channel amplitude is decoupled from
spin excitations.

It remains to calculate the triangle vertices $\hat{{\gamma }}^{\sigma _{x}}$, which are represented in a four-row-column form as $(\gamma ^{\sigma_{x}})_{i}$ and $_{j}(\gamma ^{\sigma _{x}})$. For the left vertex, one has $(\gamma ^{\sigma _{x}})_{i}\,\equiv a\hat{{\gamma }}^{\sigma _{x}}\,_{\zeta
_{2}}^{\zeta _{1}}=(1+\Gamma _{2}^{0})\langle \zeta _{1}\mathbf{p}|\sigma
_{x}/2|\zeta _{2}\mathbf{p}\rangle $ and for the right vertex,  $\,_{j}(\gamma
^{\sigma _{x}})\equiv a\,_{\zeta _{2}}^{\zeta _{1}}~\hat{{\gamma }}^{\sigma
_{x}}=(1+\Gamma _{2}^{0})\langle \zeta _{2}\mathbf{p}|\sigma _{x}/2|\zeta
_{1}\mathbf{p}\rangle$. With the use of Eq.~(\ref{eq:matrixelements}) one
obtains
\begin{align}
(\gamma ^{\sigma _{x}})_{i}\,=& \frac{(1+\Gamma _{2}^{0})}{4i}[\zeta
_{1}\delta ^{m,-1}-\zeta _{2}\delta ^{m,1}]  \label{eq:trianglei} \\
=& \frac{(1+\Gamma _{2}^{0})}{4i}\Bigg(\delta ^{m,-1}\left|
\begin{array}{c}
\psi \\
\psi%
\end{array}
\right| +\delta ^{m,1}\left|
\begin{array}{c}
-\psi \\
\psi%
\end{array}
\right| \Bigg)  \notag
\end{align}
Here the column $\psi =\left[
\begin{array}{r}
1 \\
-1%
\end{array}
\right] $ has been introduced to shorten the notations. Similarly,
\begin{align}
\,_{j}(\gamma ^{\sigma _{x}})=& -\frac{(1+\Gamma _{2}^{0})}{4i}[\zeta
_{1}\delta ^{m,-1}-\zeta _{2}\delta ^{m,1}]  \label{eq:trianglej} \\
=& -\frac{(1+\Gamma _{2}^{0})}{4i}\Bigg(\delta ^{m,-1}\left|
\begin{array}{c}
\psi \\
\psi%
\end{array}
\right| +\delta ^{m,1}\left|
\begin{array}{c}
-\psi \\
\psi%
\end{array}
\right| \Bigg)\,.  \notag
\end{align}

To proceed further, we discuss a $RA$ section in Fig.~\ref{fig:flladder}. Since we are studying the dynamic susceptibility, $\chi_{xx}(q=0,\omega )$,  the momenta in the product of the two Green's functions
$G^{R}G^{A}$ coincide. After the integration over $\xi =p^{2}/2m^{\ast }-\mu$ the product of two Green's functions yields
\begin{equation}
(G^{R}G^{A})\,_{\zeta _{2}}^{\zeta _{1}}=\frac{1}{2}\nu (\epsilon _{F})\frac{%
2\pi i}{i\omega _{n}-\Delta (\zeta _{1}-\zeta _{2})/2}\;.
\label{eq:productRA}
\end{equation}%
Here the product of the Green's functions with different chirality acquires
the difference of the frequency and the spin splitting of the energy
spectrum in the denominator, while $(GG)_{+}^{+}$ and $(GG)_{-}^{-}$ are
insensitive to the SO spin splitting. To describe the rescattering of a pair
of quasiparticles, one has to consider a ladder of $RA$ sections with the
amplitudes $\hat{\Gamma}^{k}$ in between. The amplitudes $\hat{\Gamma}^{k}$
are accompanied by the frequency summation. Ultimately, the geometrical
series of the ladder diagrams for the two-particle propagation function yields $[(G^{R}G^{A})^{-1}-(\omega _{n}/2\pi )\hat{\Gamma}^{k}]^{-1}$, where $\hat{\Gamma}^{k}$ is determined in Eqs.~(\ref{eq:gammamatrix1}) and (\ref{eq:gammamatrix2}), whereas the product $(G^{R}G^{A})$ is considered as a matrix with the diagonal elements only that
are given by Eq.~(\ref{eq:productRA}). Owing to the chiral nature of the
spectrum of the excitations the triangle vertices $\hat{{\gamma }}^{\sigma
_{x}}$ activate the channels with $m=\pm 1$ only. This fact has been already
observed in the phenomenological treatment and here reveals itself in
Eqs.~( \ref{eq:trianglei}) and (\ref{eq:trianglej}) through the Kronecker's $\delta ^{m,\pm 1}$. As a consequence, in the calculation of correlation
function $\chi _{xx}$ only the matrices $\hat{\Gamma}_{1}^{m=\pm 1}$ and $\hat{\Gamma}_{2}^{m=\pm 1}$are involved.
 
Performing the necessary matrix multiplications and the remaining frequency
summation one gets for the dynamical part of the spin correlation function
\begin{align}
&\chi _{xx}^{\text{dynamic}}(q=0,\omega )\notag\\
&\;\;=\frac{1}{8}\nu (\epsilon _{F})(1+\Gamma_{2}^{0})^{2}  \label{eq:dynamic} \\
&\quad\times \frac{-\Delta ^{2}+2\omega ^{2}(1+\Gamma _{2}^{1})(1+\Gamma
_{2}^{2})}{\Delta ^{2}\left[1+(\Gamma _{2}^{0}+\Gamma _{2}^{2})/2
\right] -\omega ^{2}(1+\Gamma _{2}^{0})(1+\Gamma _{2}^{1})(1+\Gamma
_{2}^{2}) }\;.  \notag
\end{align}
Here we performed an analytic continuation from the positive frequencies on
the Matsubara axis to the real frequency axis by $i\omega _{n}\rightarrow
\omega $. Together with the static part of $\chi _{xx}^{\text{static}}(\omega =0)=$
$(1/4)\nu (\epsilon _{F})(1+\Gamma _{2}^{0})$ this leads to the final result
\begin{align}
&\chi _{xx}^{\text{total}}(q=0,\omega )\notag\\
&\;\;=\frac{1}{8}\nu (\epsilon _{F})(1+\Gamma_{2}^{0})\label{eq:total} \\
&\quad\times \frac{\Delta ^{2}(1+\Gamma _{2}^{2})}{\Delta ^{2}\left[1+(\Gamma _{2}^{0}+\Gamma _{2}^{2})/2\right] -\omega ^{2}(1+\Gamma
_{2}^{0})(1+\Gamma _{2}^{1})(1+\Gamma _{2}^{2})}\;.  \notag
\end{align}
Remarkably, this expression reduces to Eq.~(\ref{eq:spinsusc}) obtained
phenomenologically
\begin{equation}
\chi _{xx}(q=0,\omega \rightarrow 0)_{\alpha \neq 0}=\frac{1}{8}\frac{\nu
(\epsilon _{F})}{1+\frac{1}{2}[G^{0}+G^{2}]}\;,
\end{equation}
when the relation
\begin{equation}
(1+G^{m})=1/(1+\Gamma _{2}^{m})
\end{equation}
is applied [see Chap.~2, \S~18 in Ref.~\onlinecite{Pitaevskii} and notice that in
Eqs.~(18.7) and (18.9) of the textbook $C=-\Gamma _{2}$ and $B=-\Gamma_{2}+2\Gamma _{1}$]. The resonance frequency obtained in Eqs.~(\ref{eq:dynamic}) and (\ref{eq:total}) reproduces correctly the frequency of the CSR as given by Eq.~(\ref{eq:resfreq}).

\section{Renormalizations of  disorder-induced resonance broadening and
spin-relaxation rate}
\label{sec:resonancebroadening}
In this section we first extend the treatment of the dynamic spin susceptibility to
include the disorder. This provides us with a source of the spin relaxation,
which leads to the broadening of the chiral spin resonance. Next we consider
the coupling of the spin degrees of freedom to the electromagnetic field
through the current operator. (For the sake of brevity the electric-dipole
interaction as a driving force of the spin resonance was ignored in
Sec.~ \ref{sec:flphenomenology} where the coupling via the magnetic
moment only was considered.) We show, however, that in excitation of the CSR
the electric-dipole interaction is, by far, dominating. Eventually, we compare
the intensity of the dissipation through the resonant transitions with the
nonresonant ac-Drude losses.

To account for the impurities, the RA sections in the above calculations of
the spin susceptibility should be replaced by the diffusion ladders. After
the standard averaging over the impurities the effective scattering
amplitude due to disorder is equal to $\hat{{\Xi }}=n_{\text{imp}}u^{2}(\theta _{\mathbf{p}\mathbf{p}^{\prime }})\delta _{\alpha _{1}\alpha _{2}}\delta
_{\alpha _{3}\alpha _{4}}$, where $n_{\text{imp}}$ is the number of impurities per
unit square, $u(\theta _{\mathbf{p}\mathbf{p}^{\prime }})$ is the matrix
element of the impurity potential, and the Kronecker's symbols describe the spin structure of $\hat{\Xi}$. It is assumed that for electrons at the
Fermi energy $u(\theta _{\mathbf{p}\mathbf{p}^{\prime }})$ is a function of
the scattering angle only. The disorder averaged Green's functions are
\begin{equation}
G_{\zeta }^{RA}(i\epsilon ,\mathbf{p})=\frac{1}{i\epsilon -\epsilon (
\mathbf{p})-\zeta \Delta/2\pm i/2\tau}\,,  \label{eq:Greenaver}
\end{equation}
where the scattering rate $1/\tau =\pi \nu (\epsilon
_{F})n_{imp}\left\langle u^{2}(\theta _{\mathbf{p}\mathbf{p}^{\prime
}})\right\rangle _{\theta }$. For weak enough SO\ interaction the scattering
rate $1/\tau $ is independent of the chirality $\zeta $. Note also that
the static amplitude $\hat{\Gamma}^{k}$ can be taken ignoring the influence
of the disorder when $1/\epsilon _{F}\tau \ll 1$ based on the arguments
presented in the beginning of  Sec.~\ref{sec:flphenomenology}.

To study the spin-density correlation function we sum the ladder diagrams
describing the two-particle propagation function in the electron-hole
channel. For a clean system this propagation function is equal to $
[(G^{R}G^{A})^{-1}-(\omega _{n}/2\pi )\hat{\Gamma}^{k}]^{-1}$~. Now the
multiple rescattering induced by the impurity amplitude $\hat{{\Xi }}$ and
by the \emph{e-e} interaction amplitude $\hat{\Gamma}^{k}$ should be
considered simultaneously. The impurity amplitude $\hat{{\Xi }},$ unlike $
\hat{\Gamma}^{k},$ preserves the frequency of the scattered electrons, and
therefore in the two-particles diagrams it is not accompanied by the
frequency summation. Hence, to include the impurity scattering $\hat{{\Xi }}$
in the two-particles propagation function it suffices just to modify the
previous result as follows: $[(G^{R}G^{A})^{-1}-\hat{{\Xi }}-(\omega
_{n}/2\pi )\hat{\Gamma}^{k}]^{-1}.$ As a result, for the total spin-spin
correlation function one obtains [compare with Eq.~(\ref{eq:total})]
\begin{widetext}
\begin{align}
 \chi _{xx}^{total}(q=0,\omega )=\frac{1}{8}\nu (\epsilon _{F})(1+\Gamma
_{2}^{0})    \;\;\;
 \frac{\quad \quad \Delta ^{2}(x_{2}\omega _{n}+1/\tau _{2})}{\Delta
^{2}[(x_{0}+x_{2})\omega _{n}/2+1/2\tau _{2}]+x_{0}\omega
_{n}(x_{1}\omega _{n}+1/\tau _{1})(x_{2}\omega _{n}+1/\tau
_{2})}\;.
\label{eq:totalimp}\end{align}
\end{widetext}Here the scattering rates $1/\tau _{1}$ and $1/\tau _{2}$ are
determined by the impurity scattering potential as follows: $1/\tau _{m}=\pi
\nu (\epsilon _{F})n_{imp}\left\langle (1-\exp (-im\theta ))u^{2}(\theta _{%
\mathbf{p}\mathbf{p}^{\prime }})\right\rangle _{\theta }$ with $m=1,2,$
while the frequency renormalization factors $x_{0,1,2}=1+\Gamma _{2}^{0,1,2}$. Since it is $\Gamma _{2}$ which controls the interaction in the
spin-density channel, the correlation function~(\ref{eq:totalimp}) is
determined only by the coefficients of the angular expansion of this
amplitude.

Equation~(\ref{eq:totalimp}) reveals the existence of the CSR when the
system is clean enough, $\Delta \gg 1/\tau _{1,2}.$ To determine the
position and width of the resonance, one has to perform the analytical
continuation of the retarded correlation function~(\ref{eq:totalimp}) from
the Matsubara axis to the real one and to find the roots of the cubic
polynomial in the denominator of Eq.~(\ref{eq:totalimp}). In the vicinity of
the resonance, the spin-density correlation function can be written as
\begin{equation}
\chi _{xx}^{\text{total}}(q=0,\omega )=-\frac{1}{8}\nu (\epsilon _{F})\frac{
(1+\Gamma _{2}^{0})}{(1+x_{2}/x_{0})}\;\frac{\omega _{\text{res}}^{2}}{\omega
^{2}-\omega _{\text{res}}^{2}+2i\eta _{2}\omega }\,.  \label{eq:resonance}
\end{equation}
The renormalized frequency and the width of the CSR are, correspondingly,
\begin{equation}
\omega _{\text{res}}=\Delta \left( \frac{1+x_{2}/x_{0}}{2x_{1}x_{2}}\right) ^{1/2},
\label{eq:resfr}
\end{equation}
and
\begin{equation}
\eta _{2}=\frac{1}{2x_{2}(1+x_{2}/x_{0})\tau _{2}}+\frac{1}{2x_{1}\tau _{1}}
\,.  \label{eq:reswidth}
\end{equation}
Under the condition of stability of the electron liquid all the parameters $
x_{0,1,2}$ are positive (no Pomeranchuk's instabilities) and $\eta >0$. The
positive sign of $\eta _{2}$ corresponds to the attenuation of the
spin-density excitations as it should be.

The third pole, which is purely imaginary, $\eta _{1}=-i/[(x_{0}+x_{2})\tau
_{2}],$ describes the relaxation rate of the ''chiral magnetization''. Both
relaxation rates $\eta _{1,2}$ are determined by the combinations of the
scattering rates $1/\tau _{1,2}$ only, which is natural for $\Delta \gg
1/\tau _{1,2}$.

Note that the structure of the denominator in Eq.~(\ref{eq:totalimp}) is
rich enough that regimes with other relaxation rates and different
parameters of the resonance are possible when $1/(x_{1}\tau _{1}),\;1/(x_{2}\tau _{2})$, and $1/[(x_{0}+x_{2})\tau _{2}]$ differ significantly
from each other. This is likely to happen near an instability when one of $%
x_{0,1,2}\gg 1$.

In the limit when the SO\ interaction is small (i.e., when $\Delta \ll
1/\tau _{1,2})$, it follows from Eq.~(\ref{eq:totalimp}) that the rate of
the spin relaxation is determined by the D'yakonov-Perel mechanism \cite{DyakonovP71} with a proper Fermi-liquid renormalization
\begin{eqnarray}
\chi _{xx}^{total}(q &=&0,\omega )=  \label{eq:dperelren} \\
&&\frac{1}{8}\nu (\epsilon _{F})(1+\Gamma _{2}^{0})\;\frac{\quad \Delta
^{2}\tau _{1}}{\Delta ^{2}\tau _{1}/2+(1+\Gamma _{2}^{0})\omega _{n}}\;.
\notag
\end{eqnarray}

Let us now discuss the mechanisms of excitation of the CSR. The peculiar
feature of the SO systems is that the single-particle current operator $\mathbf{J}$ contains spin [see Eq.~(\ref{eq:currentSO})]. Consequently, the
electric-dipole interaction $-(e/c)\mathbf{AJ}$ couples the electromagnetic
field $\mathbf{A}$ to the spin density. The electric-dipole interaction is a
much more effective way of excitation of the spin resonance compared to
coupling of the electromagnetic wave to the magnetic moments via the Zeeman
interaction. (To excite the CSR it is necessary to have an in-plane
component of the electric field of the radiation. This can be achieved
either in the Faraday geometry when the electromagnetic wave is incident
along the direction perpendicular to the plane of the 2DEG or in the
extraordinary Voigt geometry when the electromagnetic wave propagates
parallel to the plane of the 2DEG with the in-plane electric field.) To
clarify this issue, consider the electromagnetic wave in the Faraday
geometry with $\mathbf{E}(t)=\hat{x}E_{0}e^{ikz-i\omega t}$ and $\mathbf{B}%
(t)=\hat{y}E_{0}e^{ikz-i\omega t}.$ The energy dissipation rate according to
the Kubo formalism is determined by
\begin{equation}
Q=2\,\mathrm{Re}\,\sigma _{xx}(\omega )E_{0}^{2}+2\omega \,(g\mu _{B})^{2}\,
\mathrm{Im}\,\chi_{yy}\,(\omega )E_{0}^{2}  \label{eq:diss}
\end{equation}%
(Note that at the absence of the external magnetic field there is no
superposition contribution from the electric and magnetic-dipole
interactions to $Q$.) For the purpose of comparison of the two mechanisms,
let us confine $\sigma _{xx}(\omega )$ to the contribution originating from
spin transitions. When only the spin-term of the current operator is kept in
the correlation function determining the conductivity, it follows
immediately that $\mathrm{Re}\,\sigma _{xx}(\omega )=\,(4e^{2}\alpha
^{2}/\omega )\,\mathrm{Im}\,\chi _{yy}(\omega )$ and the dissipation rate $Q$
can be written as
\begin{equation}
Q=2\omega\, \mathrm{Im}\chi _{yy}(\omega )\Bigg[\frac{4e^{2}\alpha ^{2}}{\omega
^{2}}+(g\mu _{B})^{2}\Bigg]E_{0}^{2}\,.  \label{eq:dissspin}
\end{equation}%
For $\omega =\omega _{res}$ the first term in the square brackets does not
depend on the SO-coupling constant $\alpha $. Omitting all renormalizations,
it is $\sim e^{2}\hbar ^{2}/p_{F}^{2}$, while the second term can be
estimated as $\sim e^{2}\lambdabar^{2}$, where the Compton's length $%
\lambdabar=\hbar /m_{e}c.$ The dipole moment corresponding to the wavelength of the electrons is few orders of magnitude larger than the dipole
moment corresponding to the Compton's length, and therefore only the
electric-dipole mechanism is relevant for the excitation of the CSR. The
relative strength of the two mechanisms is $(\lambdabar/\lambda )^{2}\sim
10^{-9}-10^{-8}$.

Ignoring the momentum part of the current operator in the correlation
function that determines $\sigma _{xx}(\omega )$ is justified in the clean
system only, i.e., when the total momentum is conserved. In the presence of
disorder the situation is more subtle. Namely, in the limit $\omega \ll
1/\tau $ the momentum part of the current operator participates equally in
the excitation of the spin-flip transitions. Moreover, there are claims that
in the static limit the spin-flip transitions cannot be excited through the
electric-dipole interaction, $-(e/c)\mathbf{AJ}$, because there is a
complete cancellation between the two terms of the current operator. (This
cancellation has been noticed in Refs.~\cite{Inoue2004,Mishchenko2004,Khaetskii2004,Raimondi2004,Rashba2004} in the context
of vanishing of the static spin-Hall conductivity in the bulk of a
macroscopic system.) In the high-frequency limit the balance between
different terms of the current operator is changed, and participation of the
momentum part of the current operator in the excitation of the spin-flip
transitions becomes insignificant.

Let us clarify the action of the different terms of the current on the
spin-flip transitions when the frequency is finite. Suppose that the
current-current correlation function begins with the momentum part of the
current operator. Naively it cannot excite the spin transitions because $Tr\int d\xi \mathbf{p}_{F}\hat{\mathcal{G}}^{R}(i\epsilon ,\mathbf{p})\sigma
_{x}\hat{\mathcal{G}}^{A}(i\epsilon ,\mathbf{p})\varpropto \int d\xi
f_{\text{odd}}(\xi )$, where $f_{\text{odd}}(\xi )$ is an odd function of $\xi $. To get a
nonvanishing contribution to the spin transitions from these terms one has
to keep the dependence on $\xi $ either in the current vertex or in the
spin-splitting of the energy spectrum (both depend explicitly on the
momentum). This will inevitably be accompanied by the appearance of the
small parameter $\alpha /v_{F}$. However, the spin part of the current
contains the same parameter because it also originates from the SO
interaction. Together the two terms in the current operator give a frequency-dependent factor $g_{A\rightarrow S}=(\alpha /v_{F})[1-1/(\omega _{n}\tau+1)]$ for the effective coupling of the electromagnetic field $\mathbf{A}$
to spins (here, for the sake of brevity, impurities are assumed to be
pointlike). At low frequencies this factor approaches zero making the
excitation of the spin-flip transitions problematic. At finite frequencies
the second term in $g_{A\rightarrow S}$ originating from the momentum part
of the current operator decreases, resulting in a finite $g_{A\rightarrow S}$. This is why in the Introduction we have pointed out that it is worthwhile
to turn to AC phenomena for studying the effects of the SO interaction.
Note that the CSR is a high-frequency phenomenon. For the CSR to be narrow
enough the resonance frequency $\omega _{\text{res}}$ should much exceed $\eta _{2}\sim 1/\tau_{1,2}$. In this limit the factor $[1-1/(\omega
_{n}\tau +1)]$ in $g_{A\rightarrow S}$ approaches 1.

For completeness let us discuss the spin-Hall conductivity in the static
limit. As it has been pointed out in Ref.~\onlinecite{Mishchenko2004}, the
statement that the spin-Hall conductivity vanishes is valid only inside the
bulk of a macroscopic system. Namely, the cancellation between the two terms of the current operator has been demonstrated for $\varsigma _{xy}^{z}(q=0;\omega \ll 1/\tau )$, i.e., in a system of infinite
size. Still, in a finite-size system the spin-Hall phenomenon can exist as
the vanishing of the factor $g_{A\rightarrow S}$ may not work near the
edges. In the latter case, in a broad macroscopic system only a small
fraction of the longitudinal current that flows within a narrow strip near
the edges is effective for the spin-Hall voltage as the spin-Hall
conductivity degrades inside the bulk of the sample. For disconnected (or
weakly tunneling) edges the existence of a non-zero spin conductivity
results in the accumulation of a $z$-component of spin density at the edges.
In this connection, let us indicate that the
spin-Hall effect reported in Ref.~\onlinecite{Wunderlich04} has been observed just
at the edges of the conducting channel.

\section{Concluding Discussion}
\label{sec:Summary}
The analysis of the equations of motion performed along the lines of the
argumentation of the Kohn's theorem reveals an inherent relationship between
a transverse conductivity and a corresponding resonance in a clean system.
The same correlation function that describes the resonance determines the
value of the transverse conductivity, including its static limit. Such
relationship is useful for understanding the properties of the transverse
conductivity. For example, in Sec.~\ref{sec:eeinteraction} we demonstrate
that the absence of the \emph{e-e }renormalizations to the Hall coefficient
in a clean system is a direct consequence of the Kohn's theorem for the
frequency of the cyclotron resonance. With this in mind, in Sec.~\ref{sec:flphenomenology} we find the connection between the spin-Hall
conductivity and a spin-resonance in a 2DEG with the SO interaction. Since
this spin resonance occurs as a result of the transitions between the electron
states of different chirality which are split by the SO interaction, it is
called in this paper a chiral spin resonance.

Recently, considerable efforts have been made to determine the value of the
SO splitting in semiconductor heterostructures from the measurements of the
magnetoresistance.\cite{Nittamagnres2002,Marcus2003,Meier2004}
Another standard method for measuring the SO splitting in the electron
energy spectrum in 2DEG is the analysis of the positions of the nodes in the
beating pattern of the Shubnikov-de Haas (SdH) oscillations.\cite{Stiles88,Dasbeating89,Nitta1997,Engels1997,Sato2001} This
method, however, has certain reservations.\cite{Nittamagnres2002} In
particular, the SdH oscillations are controlled by the single-particle
relaxation time $\tau$, which in heterostructures is significantly shorter
than the transport time. In Sec.~\ref{sec:resonancebroadening} we show
that the width of the CSR can be much smaller than $1/\tau $ as it is
controlled by the scattering rates $1/\tau _{1,2}$. The CSR, if observed,
can be a useful tool for an accurate measurement of the strength of the SO
interaction.

Let us discuss the questions of the excitation and detection of the CSR. The
CSR is a limiting case of the combined resonance\cite{Rashba60} when a
static magnetic field is absent. The peculiar feature of the combined ESR
(including the CSR) is that it can be excited by the electric-dipole
interaction $-(e/c)\mathbf{AJ}$ rather than by coupling of the
electromagnetic wave to the magnetic moments via the Zeeman interaction.\cite{RashbaSheka91} The possibility of the electric-dipole excitation of the
resonance makes the observation the combined ESR a feasible task even in 2D
systems.

A problematic point in detecting of a spin resonance in 2D systems is that
a number of electrons available for spin transitions is small. A standard
method to overcome this difficulty is detecting the ESR by the
microwave-induced change of the magnetoresistivity. The resonance frequency
measured in this way\cite{Stein83,Stormer83} when extrapolated to
zero magnetic field indicates the existence of an intrinsic spin splitting.
Bychkov and Rashba\cite{BychkovRashba84} attributed this splitting to the SO
interaction induced by the structure inversion asymmetry and extracted the
value of the SO coupling constant $\alpha$.

To observe a resonance a fine-tuning control over the resonance frequency is
needed. An external magnetic field used commonly in ESR experiments may 
not be welcome for this purpose. The in-plane magnetic field makes the spin
splitting anisotropic along the Fermi surface, whereas the perpendicular
magnetic field requires an interpolation of the resonance frequency to a
zero-field limit. In addition, the orbital quantization induced by the
perpendicular magnetic field rapidly leads to the quantization of the energy
levels resulting in the quantum Hall-effect regime as it took place in Refs.~\onlinecite{Stein83} and \onlinecite{Stormer83}. Perhaps, for the CSR it is preferable to
avoid the use of the magnetic field and instead to analyze the resonance by
combining the transport measurements with the spectroscopy analysis.\cite{HuHeitmann2003} Another possible solution of the tuning problem in the case
of the CSR is the gate-voltage control of the SO splitting. For GaAs it
does not look very promising as the shift of the resonance frequency is
rather small.\cite{Yablonovitch01} However, it is known that in 
In$_x$Ga$_{1-x}$As the gate voltage strongly affect  the spin splitting that
allows  the resonance frequency to vary in a broad range.\cite{Nitta1997,Engels1997,Sato2001}

It is useful to compare the energy absorption related to the resonant
spin-flip transitions with the nonresonant heating of the 2DEG (Drude
mechanism). Assuming that the microwave radiation has a narrow frequency
range compared to the width of the resonance, one can estimate the
resonant part of the losses as $\sim e^{2}\nu \alpha ^{2}\tau
_{tr}E_{0}^{2}\sim e^{2}\Upsilon E_{0}^{2}$. At $\omega \approx \omega
_{\text{res}}\sim \Delta$, the Drude part of the dissipation is $\sim e^{2}\nu
v_{F}^{2}\tau _{tr}/(\Delta \tau _{tr})^{2}E_{0}^{2}\sim e^{2}\Upsilon
^{-1}E_{0}^{2}$. The dimensionless parameter $\Upsilon =(\Delta \tau
_{tr})(\Delta /\epsilon _{F})$ is a product of two competing factors. The
factor $\Delta /\epsilon _{F}$ characterizes the relative strength of the SO
interaction, and it is relatively small, whereas the quality factor $\Delta
\tau _{tr}\gg 1$. The ratio of the two contributions to the energy
absorption is $\thicksim $ $\Upsilon ^{2}$ .

The characteristics of the 2DEG and the data about the SO splitting are
presented in Table~\ref{tab:timeisripe}. There, we assume that the mobility $\mu _{e}$ of In$_x$Ga$_{1-x}$As is about $2\times 10^{5}\text{cm}^{2}/\text{Vs}$, which is available for
the present samples. For GaAs we take $\mu_{e}=20\times 10^{6}\text{cm}^{2}/\text{Vs}$
available only for the best reported samples. The value of the SO splitting
for In$_x$Ga$_{1-x}$As is taken from Ref.~\onlinecite{Nitta1997}, where it was
extracted from the beat pattern of SdH oscillations. For GaAs the
experimental scale of the SO interaction, $\sim 100\,\text{ns}^{-1}$, is taken from
Fig.~3 of Ref.~\onlinecite{Marcus2003}. We see that the resonance frequency in In$_x$Ga$_{1-x}$As corresponds to the far-infrared range, whereas in GaAs the
relevant frequencies are in the millimeter wave range.
\begin{table}
\begin{equation*}
\begin{array}{lccccc}
\hline\hline
\;&\;                                                            & \epsilon _{F}&\Delta &\hbar /\tau
_{tr} & \; \\
\; & n\times 10^{11}\text{cm}^{-2} & (\text{meV}) & (\text{meV}) & (\text{meV}) & \Upsilon \\ 
\hline
\text{InGaAs} & 20 & 100 & 5 & 0.1 & 5 \\
\text{GaAs} & 2 & 7 & 0.07 & 0.0008 & 1\\
\hline\hline
\end{array}%
\end{equation*}
\caption{Electronic properties of 2DEG with SO interaction.}
\label{tab:timeisripe}
\end{table}
To observe the CSR the spin splitting induced by the SO interaction should
be sufficiently isotropic. This may be realized in various situations (see
Appendix \ref{sec:BIA} for more details). One example is the asymmetrical
quantum well where the SO interaction of the structure inversion asymmetry
origin is dominant. Another variant is to fabricate a symmetric quantum well
with the [001]-growth direction and, in this way,  get rid of the SIA
spin-orbit interaction leaving only the SO interaction because of the lack of inversion symmetry of the host crystal (BIA).\cite{Dresselhaus55} The
last example is [111]-grown quantum well, which can be asymmetrical, where the
combined action of the SO interactions SIA and BIA results in the
isotropic spin splitting. It is generally accepted that in In$_{x}$Ga$_{1-x}$As
heterostructures the dominant SO interaction is because of the SIA.\cite%
{Nitta1997} However, this may be not the case for the GaAs
heterostructures where the BIA spin-orbit interaction is of comparable
strength to the SO interaction induced by the interface electric field.
The resulting spin splitting is anisotropic on the Fermi surface. This makes
an observation of the CSR in a [001]-grown GaAs heterostructure
problematic.

In view of the considerable progress in the quality of 2D heterostructures
it is worthwhile to extend the measurements of ESR to zero magnetic field.
This can give a direct information about the strength of the SO interaction.

\appendix
\section{calculation of spin-Hall conductivity in the absence of 
\lowercase{{\large\it{e-e}}} interaction}
\label{sec:Kohntheorem}
Let us calculate the spin-Hall conductivity $\varsigma _{xy}^{z}$ for noninteracting particles, copying the logic of the calculation of $\sigma _{xy}$
in Sec.~\ref{sec:eeinteraction}. [Equations (\ref{eq:SOhamilt})-(\ref{eq:Hchiral}) duplicate the corresponding equations in
Secs.~\ref{sec:eeinteraction} and \ref{sec:flphenomenology}. We repeat
them here to make this Appendix self-contained.] A single-particle
Hamiltonian with the Bychkov-Rashba SO interaction \cite{BychkovRashba84} is
\begin{equation}
H_{i}^{SO}=\frac{\boldsymbol{p}_{i}^{2}}{2m}+\alpha \lbrack \boldsymbol{p}%
_{i}\times \boldsymbol{\ell}]\cdot \boldsymbol{\sigma}_{i}\;,
\label{eq:SOhamilt}
\end{equation}%
where the unit vector $\boldsymbol{\ell}$ is perpendicular to the plane of
the 2DEG. As a result, the energy spectrum is split into two chiral branches $\epsilon _{p}^{\pm }=p^{2}/2m\pm \alpha p$.

In the presence of the SO interaction the single-particle current operator $%
\boldsymbol{j}_{i}$ contains a spin-dependent term
\begin{equation}
\boldsymbol{j}_{i}=\frac{\boldsymbol{p}_{i}}{m}+\alpha \lbrack %
\boldsymbol{\ell}\times \boldsymbol{\sigma}_{i}]\,.  \label{eq:jspin}
\end{equation}%
Since in the absence of a magnetic field the Hamiltonian~(\ref{eq:SOhamilt})
does not contain any coordinate dependence, the momentum part of the current
is time independent. Still, the current has dynamics as the current operator
contains spin.

We analyze the dynamics of spin in the chiral basis with the rotated Pauli
matrices $\mathbf{\tau }_{{\mathbf{p}}}^{\nu }=(\mathbf{a}_{\mathbf{p}}^{\nu
}\cdot \boldsymbol{\sigma}),$ where $\mathbf{a}_{\mathbf{p}}^{\nu }=\{%
\mathbf{a}^{1},\mathbf{a}^{2},\mathbf{a}^{3}\}=\{-\boldsymbol{\ell},\hat{%
\mathbf{p}},\hat{\mathbf{p}}\times \boldsymbol{\ell}\}$ and $\hat{\mathbf{p}}
$ stands for a unit vector in the direction of momentum ${\ \mathbf{p}}$. In
the chiral basis the free Hamiltonian~(\ref{eq:SOhamilt}) acquires the
diagonal form
\begin{equation}
H_{i}^{SO}=\frac{\mathbf{p}_{i}^{2}}{2m}+\alpha |\mathbf{p}_{i}|\tau _{%
\mathbf{p}_{i}}^{3}  \label{eq:Hchiral}
\end{equation}%
with the diagonal elements equal to $\epsilon _{p}^{\pm }$.

Any operator of the form $T_{i}^{\pm }=f(|\mathbf{p}_{i}|)\tau _{\mathbf{p}_{i}}^{\pm }$ has an equation of motion
\begin{equation}
\frac{d}{dt}T_{i}^{\pm }=i[H_{i}^{SO},T_{i}^{\pm }]=\pm i\omega
_{p_{i}}^{so}T_{i}^{\pm },  \label{eq:eqmotiontau}
\end{equation}%
where $\tau ^{\pm }$ are defined in the usual way, $\tau ^{\pm }=(\tau
^{1}\pm i\tau ^{2})/2,$ and
\begin{equation}
\omega _{p}^{so}=\epsilon _{p}^{+}-\epsilon _{p}^{-}=2\alpha p.
\label{eq:omegap}
\end{equation}%
These equations allows us to find the time dependence of the current
operators $j_{i}^{x}$ and $j_{i}^{y}$. For that, we express the current
components in terms of the $\mathbf{\tau }$-matrices
\begin{eqnarray}
j^{x} &=&\frac{p^{x}}{m}+\alpha \frac{p^{x}}{p}\tau _{\mathbf{p}}^{3}-\alpha
\frac{p^{y}}{p}\tau _{\mathbf{p}}^{2}\,,  \label{eq:currentx} \\
j^{y} &=&\frac{p^{y}}{m}+\alpha \frac{p^{y}}{p}\tau _{\mathbf{p}}^{3}+\alpha
\frac{p^{x}}{p}\tau _{\mathbf{p}}^{2}\;,  \label{eq:currenty}
\end{eqnarray}%
where the particle's index $i$ is omitted. The first two terms in the
expressions $j^{x,y}$ do not lead to the transitions between the chiral
states of $H_{i}^{SO}$ and therefore do not depend on time. The terms with $\tau ^{2}$ matrix induce the transitions between the states of the opposite
chirality separated by the energy $\pm \omega ^{so}$. In result, the time
dependence of the current component $j^{y}(t)$ is
\begin{equation}
j^{y}(t)=\frac{p^{y}}{m}+\frac{\alpha }{p}\big[p^{y}\tau _{\mathbf{p}
}^{3}+p^{x}(\tau _{\mathbf{p}}^{2}\cos \omega _{p}^{so}t+\tau ^{1}\sin
\omega _{p}^{so}t)\big],  \label{eq:Kohncurrenty}
\end{equation}
and a similar expression for $j^{x}(t).$

The oscillatory terms in Eq.~(\ref{eq:Kohncurrenty}) are analogous to the
oscillations in the cyclotron resonance that originate from the transitions
between the states with different circulation; compare Eqs.~(\ref%
{eq:eqmotionJ}) and (\ref{eq:eqmotiontau}). Furthermore, the same
description applies for the case of the ESR, where the transitions in the
external magnetic field occur between the states of the opposite spin
polarization.

We are currently in the stage when the application of the Kohn's
argumentation for the time evolution of the current operators allows us to
calculate the transverse conductivity. The transverse spin conductivity $%
\varsigma _{xy}^{z}$ describes the response of the spin-$z$-component
current in the $x$ direction, $\mathfrak{j}_{z}^{x}=\frac{1}{4}(\sigma
^{z}j^{x}+j^{x}\sigma ^{z})$, to the electric field applied in the $y$ direction. In the chiral basis $\sigma ^{z}=$ $-\tau ^{1}$, and therefore $\mathfrak{j}_{z}^{x}=-(p^{x}/2m)\tau ^{1}$. The transverse spin conductivity
is given by the corresponding Kubo formula (we restore the particles index $i$)
\begin{equation}
\varsigma _{xy}^{z}=\frac{e}{\omega }\int_{0}^{\infty }dt\,e^{i\omega
t}\left\langle \sum \left[ \mathfrak{j}_{z;i}^{x},j_{i}^{y}(-t)\right] \right\rangle \,,
\label{eq:spinKubo}
\end{equation}%
which with the use of Eq.~(\ref{eq:Kohncurrenty}) yields
\begin{equation}
\varsigma _{xy}^{z}=e\sum \left\langle\frac{\alpha }{mp_{i}}\Big[-\frac{%
p_{i}^{x}p_{i}^{y}\tau _{\mathbf{p}_{i}}^{2}}{\omega ^{2}}+\frac{\left(
p_{i}^{x}\right) ^{2}\tau _{\mathbf{p}_{i}}^{3}}{\omega ^{2}-(\omega
_{p_{i}}^{so})^{2}}\Big]\right\rangle.  \label{eq:spinKuboresult}
\end{equation}%
To get the final result one has to perform the average $\langle \cdots
\rangle $ in this equation. When averaged, the term with $\tau ^{2}$ matrix
vanishes, $\langle \tau _{\mathbf{p}}^{2}\rangle =0$, because the
spin-dependent term in Hamiltonian~(\ref{eq:SOhamilt}) contains $\tau ^{3}$
matrix only. As the chiral states are eigenstates with energies $\epsilon _{\mathbf{p}}^{\pm }$, the population of a state $\mathbf{p}$ is equal to $n_{F}(\epsilon _{\mathbf{p}}^{+})$ for the $+$ chirality state and $n_{F}(\epsilon _{\mathbf{p}}^{-})$ for the $-$ chirality state; $n_{F}(\epsilon )=[\exp (\epsilon -\mu )+1]^{-1}$. Correspondingly, the
expected value of $\tau _{\mathbf{p}}^{3}$ is equal to $\langle \tau _{%
\mathbf{p}}^{3}\rangle =n_{F}(\epsilon _{p}^{+})-n_{F}(\epsilon _{p}^{-})$.
Finally, this yields
\begin{align}
\varsigma _{xy}^{z} =&e\int \frac{d^{2}p}{(2\pi )^{2}}\frac{1}{8\alpha mp}
[n_{F}(\epsilon _{p}^{-})-n_{F}(\epsilon _{p}^{+})]  \notag \\
=&\frac{e}{8\alpha m}\int \frac{dp}{2\pi }[n_{F}(\epsilon _{{p}}^{-})-n_{F}(\epsilon _{{p}}^{+})]\,.  \label{eq:spinconductivity}
\end{align}
Unlike, the cyclotron resonance, where all electrons precess together and
contribute equally to $\sigma _{xy}=-(ne^{2}/m)\omega _{c}^{-1}$, in the
case of spin-Hall conductivity the contribution from electrons of the
opposite chirality tend to cancel each other out. The factor $1/\alpha $ in
Eq.~(\ref{eq:spinconductivity}) is equivalent to $1/\omega _{c}$ in $\sigma
_{xy}$, but due to the cancellation only a stripe of the width $2\alpha m$
between the Fermi surfaces of electrons of the opposite chirality
contributes that makes the value $\varsigma _{xy}^{z}$ finite in the limit
of small $\alpha $. For non-interacting electrons it is also possible to
express $\varsigma _{xy}^{z}$ as a contribution from the bottom of the band.
For that, rewrite $\epsilon _{p}^{\pm }$ as $\epsilon _{p}^{\pm }=\frac{1}{2m%
}(p\pm \alpha m)^{2}-\frac{1}{2}m\alpha ^{2}$ and shift the momentum
variables to $q^{\pm }=p\pm \alpha m$. Then
\begin{equation}
\varsigma _{xy}^{z}=\frac{e}{4\alpha m}\int_{0}^{\alpha m}\frac{dq}{2\pi }%
n_{F}(\frac{q^{2}}{2m}-\frac{m\alpha ^{2}}{2})\,.  \label{eq:shift}
\end{equation}%
At low temperatures $n_{F}(0)=1$ and $\varsigma _{xy}^{z}=e/8\pi $ for any
finite $\alpha .$ On the other hand, when $\alpha m\ll p_{F}$ one can get
for arbitrary temperatures $\varsigma _{xy}^{z}\approx \left( e/8\pi \right)
\,n_{F}(\epsilon =0;T)\,,$where $n_{F}(\epsilon =0;T)$ is the occupation
number at the bottom of the band at a temperature $T$.

The expression for $\varsigma _{xy}^{z}$ as given by Eq.~(\ref%
{eq:spinconductivity}) reminds the corresponding expression for the static
spin susceptibility: $\chi =(g\mu _{B}/2B)\int [d^{2}p/(2\pi)^{2}]\left[n_{F}(\epsilon _{p}-\frac{1}{2}\Delta_Z )-n_{F}(\epsilon _{p}+\frac{1}{2}\Delta_Z )\right]$, where $\Delta_Z =g\mu _{B}B$. This gives a hint why there exists
a connection\cite{Dimitrova2004,3Loss2004} between the spin-density
correlation function and $\varsigma _{xy}^{z}$. In Section~\ref{sec:eeinteraction} a direct connection between $\varsigma _{xy}^{z}$ and
the dynamic (retarded) spin susceptibility has been derived [see Eqs.~(\ref{eq:spinHalltotal})-(\ref{eq:SScor})]
\begin{eqnarray}
\varsigma _{xy}^{z} &=&\frac{e}{m}\chi _{xx}(q=0,\omega )\;,
\label{eq:relation} \\
\chi _{xx}(q &=&0,\omega )=\frac{i}{4}\int_{0}^{\infty }dt\,e^{i\omega
t}\left\langle \sum \left[ \sigma _{i}^{x}(t),\sigma _{i}^{x}(0)\right] \right\rangle\,.  \notag
\end{eqnarray}
A direct calculation of $\chi _{xx}(q=0,\omega )$ in the presence of the SO
interaction can be done straightforwardly with the help of Eq.~(\ref%
{eq:Kohncurrenty}). In the limit of small frequencies, $\omega \ll \omega
^{so},$ the correlation function $\chi _{xx}(q=0,\omega \rightarrow
0)_{\alpha \neq 0}$ is equal to a half of the static spin susceptibility of
a free 2DEG in the absence of the SO interaction
\begin{equation}
\chi _{xx}(q=0,\omega \rightarrow 0)_{\alpha \neq 0}=\frac{m}{8\pi }~.
\label{eq:spinsuscept}
\end{equation}
Correspondingly, in the presence of the SO interaction the spin-Hall
conductivity $\varsigma _{xy}^{z}=e/8\pi \,$.

\section{Fermi-liquid analysis of spin resonance in the presence of  
Dresselhaus SO interaction (BIA)}
\label{sec:BIA} 
In this appendix we analyze the kinetics of electrons in the
presence of the SO interaction of the BIA origin.\cite{Dresselhaus55} The
spin-orbit interaction in the semiconductors with the zinc-blende crystal
structure is described by the Hamiltonian
\begin{equation}
H_{\text{bulk}}^{\text{SO}}=\gamma \lbrack \sigma _{x}k_{x}(k_{y}^{2}-k_{z}^{2})+\text{c.p.}]~,
\label{eq:BIAbulk}
\end{equation}%
where c.p. stands for the cyclic permutations. For 2DEG the Hamiltonian (\ref{eq:BIAbulk}) leads to a linear in momentum term in the SO interaction.\cite{Eppenga1988} For the case of the [001]-grown quantum well the linear term
can be obtained from $H_{\text{bulk}}^{\text{SO}}$ by replacing $k_{z}^{2}$ and $k_{z}$ by
their averages $\langle k_{z}^{2}\rangle $ and $\langle k_{z}\rangle =0$
\begin{equation}
H_{[001]}^{\text{SO}}=\beta (p_{x}\sigma _{x}-p_{y}\sigma _{y}),
\label{eq:BIAlinear}
\end{equation}%
where $\beta =-\gamma \langle k_{z}^{2}\rangle $. Unlike the Rashba
Hamiltonian (\ref{eq:BRashbaSO}), this term does not have a structure of a
triple scalar product and, therefore, it is not rotationally invariant.
Nevertheless, it leads to the isotropic spin splitting of the energy
spectrum.

The structure of the linear term that is formally identical to the
Hamiltonian (\ref{eq:BRashbaSO}) can be realized in the
[111]-grown quantum well. This is a consequence of a threefold
rotation symmetry with respect to [111] axis. On the contrary, SO
interaction in the [110]-grown quantum well results in the
anisotropic spin splitting and therefore such heterostructures are
not suitable for the observation of CSR.

In the following we analyze the kinetic equation (\ref{eq:linearized}) for
the case [001]-grown 2DEG when the linear SO interaction term is given by
Eq. (\ref{eq:BIAlinear}). The remaining cubic terms in the SO interaction is
a source of the CSR broadening. It can be neglected for the narrow enough
quantum well when $\langle k_{z}^{2}\rangle \gg k_{F}^{2}$.

It is instructive to start with the Rashba and Dresselhaus interactions
acting together. The combined action of the BIA and SIA mechanisms of the
spin-orbit interaction is described by the Hamiltonian
\begin{align}
H_{\text{linear}}^{\text{SO}}=& \alpha (p_{y}\sigma _{x}-p_{x}\sigma _{y})+\beta
(p_{x}\sigma _{x}-p_{y}\sigma _{y}) \notag \\
=& \frac{1}{2}\,\Delta _{\mathbf{p}}\,[a_{\mathbf{p}}\sigma _{x}+b_{\mathbf{p}}\sigma _{y}]\,,  \label{eq:SIABIA}
\end{align}
where $a_{\mathbf{p}}^{2}+b_{\mathbf{p}}^{2}=1.$ In the presence of the two
SO interaction terms the spin splitting energy $\Delta _{\mathbf{p}}$ is a
varying function along the Fermi surface
\begin{equation}
\Delta _{\mathbf{p}}=2p_{F}\Lambda _{\mathbf{p}}\,,\quad \Lambda _{\mathbf{p}%
}=\big[\alpha ^{2}+\beta ^{2}+2\alpha \beta \sin 2\theta _{\mathbf{p}}\big]%
^{1/2}\,.  \label{eq:delatap}
\end{equation}%
The coefficients $a_{\mathbf{p}}$ and $b_{\mathbf{p}}$ are defined as
\begin{align}
a_{\mathbf{p}}=& \Lambda _{\mathbf{p}}^{-1}(\;\;\alpha \sin \theta _{\mathbf{%
p}}+\beta \cos \theta _{\mathbf{p}})  \notag  \label{eq:allomega} \\
b_{\mathbf{p}}=& \Lambda _{\mathbf{p}}^{-1}(-\alpha \cos \theta _{\mathbf{p}%
}-\beta \sin \theta _{\mathbf{p}})\,.
\end{align}%
We introduce a set of the Pauli matrices such that the SO spin splitting
term takes the form $\delta \epsilon ^{SO}=\frac{1}{2}\Delta _{\mathbf{p}%
}\,t_{\mathbf{p}}^{3}$, namely,
\begin{align}
t_{\mathbf{p}}^{1}=& -\sigma _{z}\,,\quad t_{\mathbf{p}}^{2}=-b_{\mathbf{p}%
}\sigma _{x}+a_{\mathbf{p}}\sigma _{y}\,,  \notag \\
t_{\mathbf{p}}^{3}=& a_{\mathbf{p}}\sigma _{x}+b_{\mathbf{p}}\sigma _{y}\,.
\label{eq:t-matrices}
\end{align}%
The renormalization of the $\beta$ term by the \emph{e-e} interaction is
analyzed in the same way as the renormalization of the Bychkov-Rashba
coefficient $\alpha $ in Eq.~(\ref{eq:energy}) yielding $\beta ^{\ast
}/\beta =\alpha ^{\ast }/\alpha =1/(1+G^{1})$.

The kinetic equation similar to Eqs.~(\ref{eq:u-eqa})-(\ref{eq:u-eqc}) can now be
written using the expansion of $\hat{u}(\theta _{\mathbf{p}})$ in terms of
the $t_{\mathbf{p}}$--matrices: $\hat{u}(\theta _{\mathbf{p}})=\sum {u}%
_{i}(\theta _{\mathbf{p}})t_{\mathbf{p}}^{i}$. The force term is equal to
\begin{align}
\lbrack \delta \hat{\epsilon}_{\mathbf{p}}^{SO}, \,\sigma _{x}]\,\mathfrak{F}\,e^{i\omega t}
= -it^{1}\,(2p_{F}\alpha ^{\ast }\cos \theta _{\mathbf{p}}+2p_{F}\beta
^{\ast }\sin \theta _{\mathbf{p}})\,\mathfrak{F}\,e^{i\omega t}\,,  \label{eq:BIAdrforce} 
\end{align}%
and the kinetic equation acquires the form
\begin{subequations}
\begin{align}
\Delta _{\mathbf{p}}^{-1}\frac{du_{1}(\theta _{\mathbf{p}})}{dt}=&
u_{2}(\theta _{\mathbf{p}})+\int d\theta _{\mathbf{p}^{\prime }}G(\theta _{%
\mathbf{p}\mathbf{p}^{\prime }})C^{3,2}(\theta _{\mathbf{p}},\theta _{%
\mathbf{p}^{\prime }})u_{2}(\theta _{\mathbf{p}^{\prime }})  \notag \\
-\Delta _{\mathbf{p}}^{-1}[2p_{F}& \alpha ^{\ast }\cos \theta _{\mathbf{p}%
}+2p_{F}\beta ^{\ast }\sin \theta _{\mathbf{p}}]\mathfrak{F}e^{i\omega t}\,,
\label{eq:app-u-eqa} \\
\Delta _{\mathbf{p}}^{-1}\frac{du_{2}(\theta _{\mathbf{p}})}{dt}=&
-u_{1}(\theta _{\mathbf{p}})-\int d\theta _{\mathbf{p}^{\prime }}G(\theta _{%
\mathbf{p}\mathbf{p}^{\prime }})u_{1}(\theta _{\mathbf{p}^{\prime }})\,.
\label{eq:app-u-eqb}
\end{align}%
Here the structure factor $C^{3,2}$ appears because of the commutator $[t_{%
\mathbf{p}}^{3},t_{\mathbf{p^{\prime }}}^{2}]=-2it^{1}C^{3,2}(\theta _{%
\mathbf{p}},\theta _{\mathbf{p}^{\prime }})$, where
\end{subequations}
\begin{align}
C^{3,2}& (\theta _{\mathbf{p}},\theta _{\mathbf{p}^{\prime }})=a_{\mathbf{p}%
}a_{\mathbf{p}^{\prime }}+b_{\mathbf{p}}b_{\mathbf{p}^{\prime }}
\label{eq:C32} \\
=& (\Lambda _{\mathbf{p}}\Lambda _{\mathbf{p^{\prime }}})^{-1}[(\alpha
^{2}+\beta ^{2})\cos (\theta _{\mathbf{p}}-\theta _{\mathbf{p}^{\prime
}})+2\alpha \beta \sin (\theta _{\mathbf{p}}+\theta _{\mathbf{p}^{\prime
}})]\,.  \notag
\end{align}%
The structure factor $C^{3,2}$ reduce to $\cos (\theta _{\mathbf{p}}-\theta
_{\mathbf{p}^{\prime }})$ when only one of the SO interactions (BIA or SIA)
is acting.

The kinetic equation has the same form when either the SIA or BIA
mechanism acts solely. Hence the pure BIA system exhibits the same chiral
spin resonance with the frequency given by Eq.~(\ref{eq:resfreq}) and $\Delta \rightarrow \Delta _{\text{BIA}}$. Actually this observation, as well as
equal renormalization of $\alpha $ and $\beta$, is related to a duality of
the $\alpha $- and $\beta $-SO terms. Namely, $H_{linear}^{SO}$ is symmetric
with respect to a simultaneous rotation of the Pauli matrices around the
direction $\hat{n}=(\hat{x}+\hat{y})/\sqrt{2}$ by $\pi ,$ i.e., $\sigma
_{x}\rightleftarrows \sigma _{y},\;\sigma _{z}\rightarrow -\sigma _{z}$,
together with the replacement $\alpha \rightleftarrows -\beta $, whereas the
\emph{e-e} interaction is symmetric with respect to any spin rotations.

\begin{acknowledgments}
We thank Alex Punnoose, Emmanuel Rashba, and Yehoshua Levinson for
stimulating discussions of the problems related to the spin-Hall
conductivity. We thank Jurgen Smet and Gerhard Abstreiter for a useful
discussion of the microwave experiments. The work was supported by the
US-Israel Binational Science Foundation (BSF) and the Minerva Foundation.
\end{acknowledgments}

\end{document}